%% file: main.tex
\documentclass[twocolumn, twocolappendix]{aastex701}

\hypersetup{linkcolor=red,citecolor=blue,filecolor=cyan,urlcolor=magenta}

\input{packages.tex} 
\input{macros.tex} 
\input{variables.tex} 

\begin{document}

\title{Evidence for Environmental Stripping in the Coma Cluster}

\author[orcid=0000-0003-1595-9903]{Richard T. Pomeroy}
\affiliation{Department of Physics and Astronomy, The University of Texas Rio Grande Valley, Brownsville, TX 78520, USA}
\email[show]{richard.pomeroy01@utrgv.edu}
\correspondingauthor{Richard T. Pomeroy}

\author{Juan P. Madrid}
\affiliation{Department of Physics and Astronomy, The University of Texas Rio Grande Valley, Brownsville, TX 78520, USA}
\affiliation{National Science Foundation, 2415 Eisenhower Avenue, Alexandria, VA 22314, USA}
\email{juan.madrid@utrgv.edu}

\author{Conor R. O'Neill}
\affiliation{Australian Astronomical Observatory, PO Box 915, North Ryde, NSW 1670, Australia}
\email{conor.oneill.physics@gmail.com}

\author[0000-0003-4906-8447]{Alexander T. Gagliano}
\affiliation{\IAIFI}
\affiliation{\CfA}
\affiliation{\MIT}
\email{gaglian2@mit.edu}

\begin{abstract}
The stability and longevity of globular clusters (GCs) make them effective tracers of the dynamical histories of galaxies in cluster environments. We construct a catalog of 23,351 GC candidates in the Coma cluster using imaging from the Hubble Space Telescope Advanced Camera for Surveys. We cross-match galaxy data from the SIMBAD, NED, and SDSS archives to construct a galaxy sample and model their GC populations using the GC specific frequency. We find several galaxies with significantly smaller GC populations than expected from their luminosities, consistent with either tidal stripping or intrinsically low formation efficiencies. We analyze annular and Voronoi GC radial profiles of the BCGs (NGC~4874 and NGC~4889) and other Coma galaxies. A 2D Voronoi density mapping reveals GC populations with marked deficits compared to our modeled expectations, including galaxies in proximity to the BCGs (\eg IC~3998, NGC~4875, NGC~4876) and others distributed across Coma (\eg NGC~4908, NGC~4883, IC~4042). Azimuthal symmetry testing suggests past dynamical interactions may have truncated GC systems in some galaxies, while intrinsic deficits are probable in others (\eg IC~3973, IC~3976, IC~4040, IC~4045). Our results show that GC deficits exist in several Coma galaxies and that the 2D density structure reveals environmental signatures, with asymmetry statistics consistent with directional stripping. These findings highlight GC populations as powerful probes of environmental processing and the dynamical histories of galaxies in dense cluster environments.
\end{abstract}


\par
\section{Introduction}\label{sec:intro}
It has been established that globular clusters (GCs), and the globular cluster systems (GCSs) they comprise in galaxy halos, formed predominantly during the early phases of galaxy assembly at high redshift. This view is supported both observationally \citep[\eg][]{Strader_2005,Brodie_2006,Vanzella_2022} and by simulations such as E-MOSAICS and SIEGE \citep[\eg][]{Pfeffer_2018,Creasey_2019,Horta_2021,Pascale_2025} and suggest these GCS coevolve with their host galaxies. This in turn implies a connection between the properties of the GCSs and their host galaxies \citep[\eg][]{Harris_2013} and many empirical relationships have been found. For example, there are relationships between GC population count and galaxy luminosity/mass \citep[\eg][]{Harris_1981,Peng_2008,Harris_2013}, GCS mass and galaxy halo mass \citep[\eg][]{Spitler_2009,Misgeld_2011,Hudson_2014,Harris_2017a, Hudson_2018} and also the spatial extent of GCSs \citep[\eg][]{Madrid_2018,Lim_2024,Pomeroy_2025,D'Abrusco_2025}, which scale differently with red (metal-rich, in-situ formation) and blue (metal-poor, accreted) sub-populations.

The nature of these empirical relations is potentially more complex in dense galaxy cluster environments, where repeated interactions, tidal forces, and mergers can significantly modify both the galaxies and their surrounding GCSs. Galaxy–galaxy encounters, tidal stripping by the cluster potential, and pre-processing within infalling galaxy groups prior to cluster accretion can redistribute or remove GCs from their original hosts, contributing to the intracluster GC population \citep[e.g.,][]{West_1995, Bekki_2003,Peng_2011,Madrid_2017,Forbes_2018a,Pfeffer_2023}. These environmental processes also alter the structural and kinematic properties of GCSs, steepening or truncating GCS radial profiles \citep[e.g.,][]{Smith_2013,Durrell_2014}, and can bias the observed scaling relations between GC counts, galaxy luminosity, and halo mass \citep[\eg][]{Harris_2013, Hudson_2014}. As a result, understanding the interplay between intrinsic galaxy properties and environmental effects is essential when interpreting the observed variation in GC counts and spatial distributions in massive clusters such as Coma, Virgo, and Fornax.

There are published cases where galaxies have been found to host unusually low GC populations relative to their luminosity or mass. Notable examples include NGC~1277 in the Perseus cluster \citep{Beasley_2018}, a compact relic galaxy thought to have formed early and experienced little subsequent evolution, and ultra-diffuse galaxies (UDGs) in Coma \citep{Lim_2018}. However, UDGs exhibit a wider range of GC specific frequencies than typical dwarf galaxies. In their study of 48 Coma UDGs, \citet{Lim_2018} found that some host two to three times more GCs than more luminous galaxies.

\defcitealias{Pomeroy_2025}{RTP25}
This paper builds on observations and results from an earlier paper, \citet{Pomeroy_2025}, hereafter \citetalias{Pomeroy_2025}, where we identified a lack of compact stellar systems (CSSs), \ie GCs and ultra-compact dwarfs (UCDs) in close proximity ($\lesssim\!\qty{5}{\kilo\parsec}$) to many of the brighter galaxies in the central \qty{\approx1}{\mega\parsec} of the Coma Cluster. There are also hints of several galaxies (\ie $\sim5-10$) in the region which have a disproportionately low surface density of GCs within eight effective radii (8\,\Reff), where \Reff\  denotes the galaxy effective radii, enclosing half the total stellar light. As most studies in the past have concentrated on observation and analysis of galaxies with large numbers of GCs, there is an opportunity to further explore this dataset, which covers the entire core of Coma. Consequently, we aim to test the existence of multiple low-GC population galaxies in the Coma cluster and investigate how environmental processes may bias interpretations based on GC specific frequency in a dense cluster environment.

The focus of this work, therefore, is to explore the nature of galaxies with anomalous GCS in the Coma cluster (Abell 1656) using the dataset and methods outlined in Section \ref{sec:p2_methods}. We begin the study in Section \ref{sec:gal_light_model}, by comparing the GCS around two galaxies which appear to have a low surface density of GCs. We progress with a Voronoi tessellation based analysis of the GC radial profile about the two brightest cluster galaxies (BCGs), shown in Section \ref{sec:1D_voronoi}. Following this a review of the GC specific frequencies of galaxies in Coma in Section \ref{sec:Coma_gal_GCSF}. We then expand on the Voronoi analysis in Section \ref{sec:2D_voronoi_density}, by developing a 2D density map of GCS throughout the entire observed central region of Coma, following up in detail on regions exhibiting low GCS density. We look for signs of interaction in the cluster environment in Section \ref{sec:az_symmetry}, by exploring the azimuthal symmetry on these and other galaxies in our sample. In Section \ref{sec:p2_results_discuss}, we discuss the result of these analyses before concluding in Section \ref{sec:p2_conclusion}.

A distance to Coma of \qty{100}{\mega\parsec} ($(m-M)\qty{=35.0}{\mag}$) is adopted \citep{Carter_2008}.

\section{Data and Methods}\label{sec:p2_methods}

The original GC dataset was presented in \citet{Madrid_2018}, with refinements detailed in \citetalias{Pomeroy_2025}. The final catalog used here comprises \RTPVarCSSTotal\ CSS candidates, including \RTPVarGCTotal\ GC candidates and \RTPVarUCDTotal\ UCD candidates.

The final CSS catalog used here includes both GC and UCD candidates, following the classification scheme described in \citetalias{Pomeroy_2025}. Since the sample is overwhelmingly dominated by GC candidates, we analyze the full CSS catalog together in what follows, and distinguish explicitly between GCs and UCDs only where required.

\begin{figure*}[!p]
    \centering
    \includegraphics[width=0.76\linewidth]{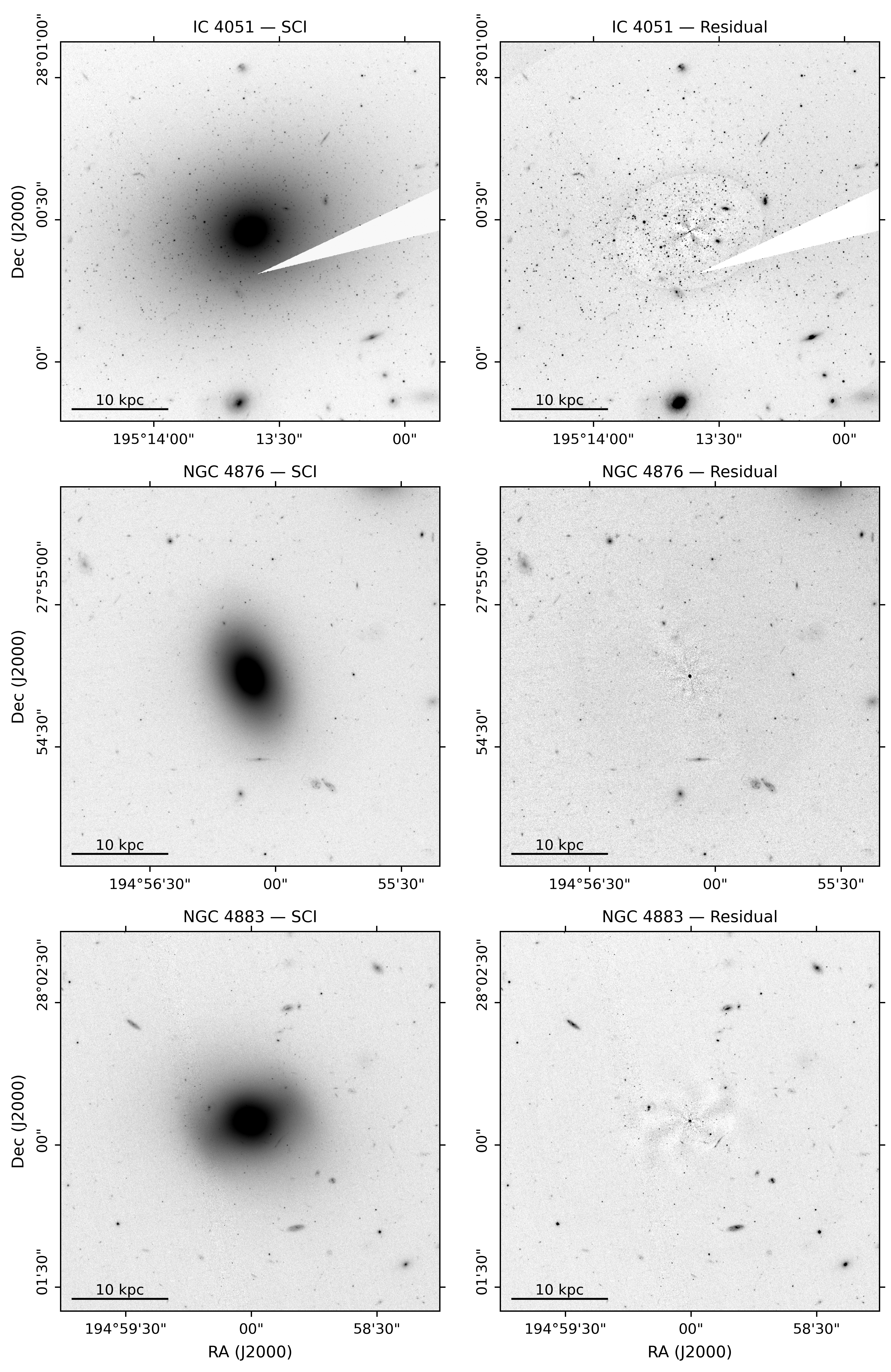}
    \caption{HST F814W science (left panels) and residuals (right panels) after subtraction of galaxy light, modeled with elliptical isophotes as described in Section \ref{sec:gal_light_model}. Residuals at galaxy centers are still visible at this contrast level, but the GCS comparison between the galaxies is of note. The deficit of GCs around galaxies NGC~4876 (middle, $\MV=\qty{-20.51}{\mag}$) and NGC~4883 (lower, $\MV=\qty{-20.66}{\mag}$), identified in Section \ref{sec:Coma_gal_GCSF}, compared to the excess around IC~4051 (top, $\MV=\qty{-21.66}{\mag}$), is clear.}
    \label{fig:sci_res_galaxies}
\end{figure*}

The ACS data used for this work were obtained under HST programs: GO 10861 \citep{Carter_2008}, GO 11711 \citep{Cho_2016} and GO 12918 \citep{Harris_2017b} with the reduction process given in \citet{Madrid_2018}. The ACS pointings cover the core ($\sim1$Mpc) of Coma, including NGC 4889 and NGC 4874, the two central dominant galaxies that define Coma’s unusual dual-BCG system. The ACS data also include IC~4051 a giant elliptical galaxy that has a large population of GCs \citep{Woodworth_2000, Madrid_2018}, and UCDs \citepalias{Pomeroy_2025}. The ACS data were obtained using two filters: F475W and F814W. 

\subsection{Galaxy sample, cross-matching, and derived quantities}
\label{ssec:gal_sample}

We constructed an initial galaxy sample in the Coma region by querying the SIMBAD TAP service within a fixed sky window (RA $\ang{\sim194.769;;}$ to $\ang{\sim195.289;;}$, Dec $\ang{\sim27.826;;}$ to $\ang{\sim28.122;;}$). Sources were selected with SIMBAD object type \texttt{G..} (galaxies), redshifts $0.012 \lesssim z \lesssim 0.0441$, and apparent magnitudes $\MV<19$~mag, yielding an initial sample of 173 systems. The magnitude cut was affected to isolate the luminous galaxy population relevant to the GC-system modeling while avoiding poorly constrained low-luminosity systems. SIMBAD coordinates, redshifts, and literature-based photometry were used solely for sample identification and filtering.

For each galaxy, we retrieved available morphological dimensions from SIMBAD, including major and minor axes and position angles at the $B=25$~mag~arcsec$^{-2}$ isophote (\texttt{galdim\_majaxis}, \texttt{galdim\_minaxis}, \texttt{galdim\_angle}), together with provenance metadata. When SIMBAD size measurements were unavailable, corresponding values were supplemented from NED diameter catalogs.

The primary quantitative galaxy properties used in this work were derived from homogeneous SDSS DR17 photometry \citep{Abdurrouf_2022}. All sources were cross-matched to SDSS within $3\arcsec$, retaining primary galaxy detections (\texttt{mode}=1, \texttt{type}=3). From SDSS we extracted extinction-corrected $g$, $r$, and $i$ cModel magnitudes and derived apparent $V$-band magnitudes using standard $g-r$ transformations. These were adopted as the principal basis for computing absolute magnitudes, providing a uniform luminosity scale across the sample.

Galaxy effective radii were likewise obtained preferentially from SDSS $r$-band model fits. We adopted \texttt{deVRad\_r} or \texttt{expRad\_r} according to the fractional de~Vaucouleurs contribution (\texttt{fracDeV\_r}), and constructed both semi-major-axis and circularized estimates using the fitted axis ratios (\texttt{Re\_r\_arcsec}, \texttt{Re\_r\_arcsec\_circ}). These SDSS-based measurements define the primary, homogeneous size scale used throughout the analysis. For the small subset of galaxies without SDSS photometric matches (11/173), we adopt a proxy $\Reff \approx \psub{D}{maj}/8$ based on catalog major-axis diameters and flag these systems accordingly.
 
\section{Galaxy light modeling}\label{sec:gal_light_model}

To illustrate the deficit in GCSs is a real effect, we use two galaxies from \citetalias{Pomeroy_2025} which we suspect have a low GC population, and produce a model of the galaxy light which we can subtract, to show the GCS. 

Using the \texttt{isophote} module from the \texttt{photutils} package, we first create a list of isophotes using the python class \texttt{IsophoteList}, and after validating these visually as regions in \texttt{DS9}, a galaxy light model is constructed using \texttt{build\_ellipse\_model}. The original HST F814W SCI image and model subtracted residual for NGC~4876 and NGC~4883 are shown in the middle and lower image pairs of \cref{fig:sci_res_galaxies}, respectively. We carry out the same process on IC~4051, with its extensive excess in GCS \citep{Woodworth_2000, Madrid_2018} for comparative purposes, included as the top pair of images in \cref{fig:sci_res_galaxies}. 

These three galaxies, differ in visual magnitude by \qty{\sim1.0}{\mag} and thus, in the context of GCS comparison, have broadly similar magnitudes. Despite this, their GC populations are visually distinct, with IC~4051 showing a much greater surface density of GCs than either NGC~4876 or NGC~4883. 

We take this observation as motivation to investigate these galaxies further, along with other galaxies in the Coma cluster.

\begin{figure*}[!p]
\begin{subfigure}{0.49\linewidth}
    \centering
    \includegraphics[width=1\linewidth]{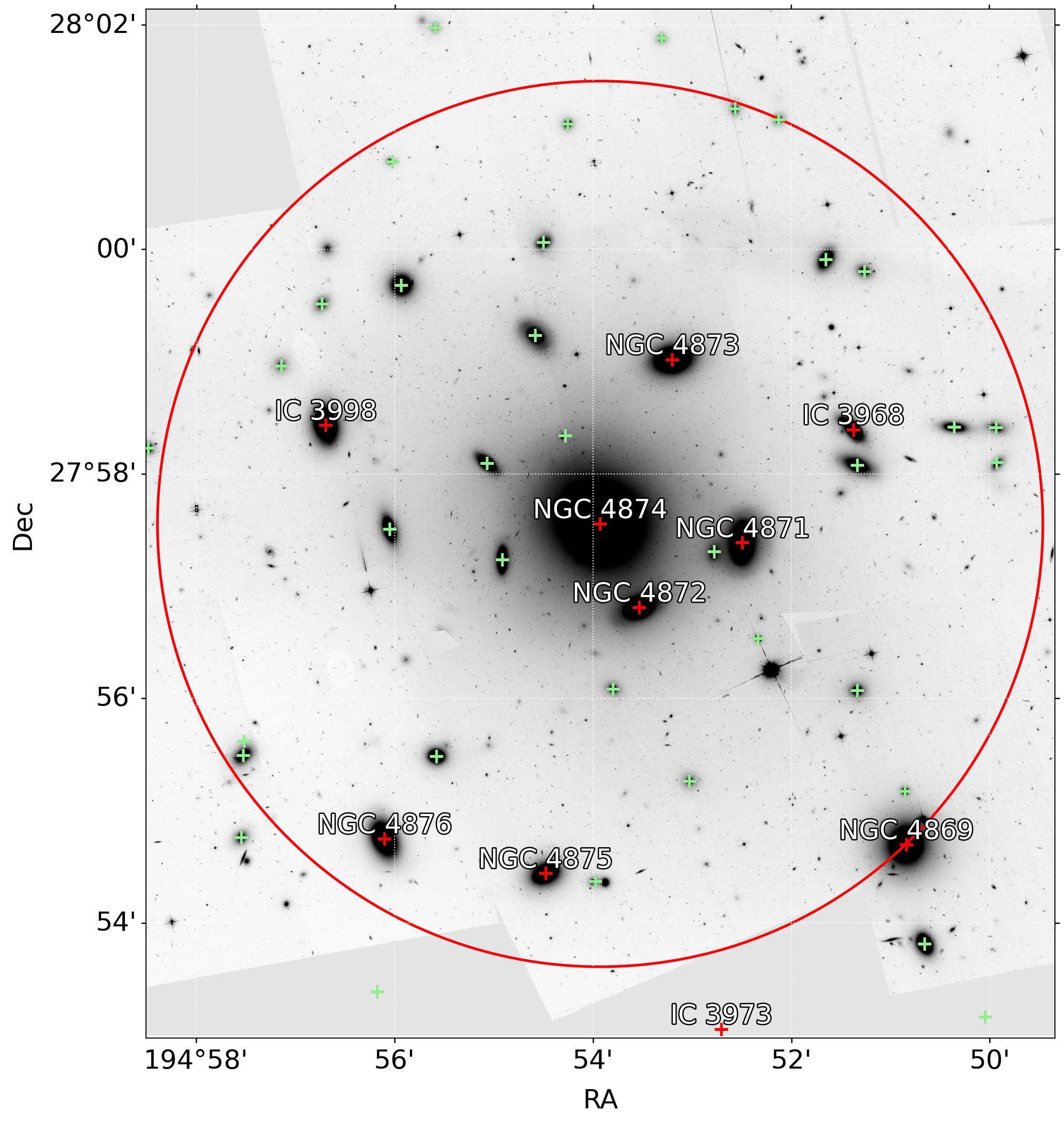}
    \caption{HST F814W image of NGC~4874 region.}
    \label{fig:Coma_NGC 4874_hst}
\end{subfigure}
\begin{subfigure}{0.49\linewidth}
    \centering
    \includegraphics[width=1\linewidth]{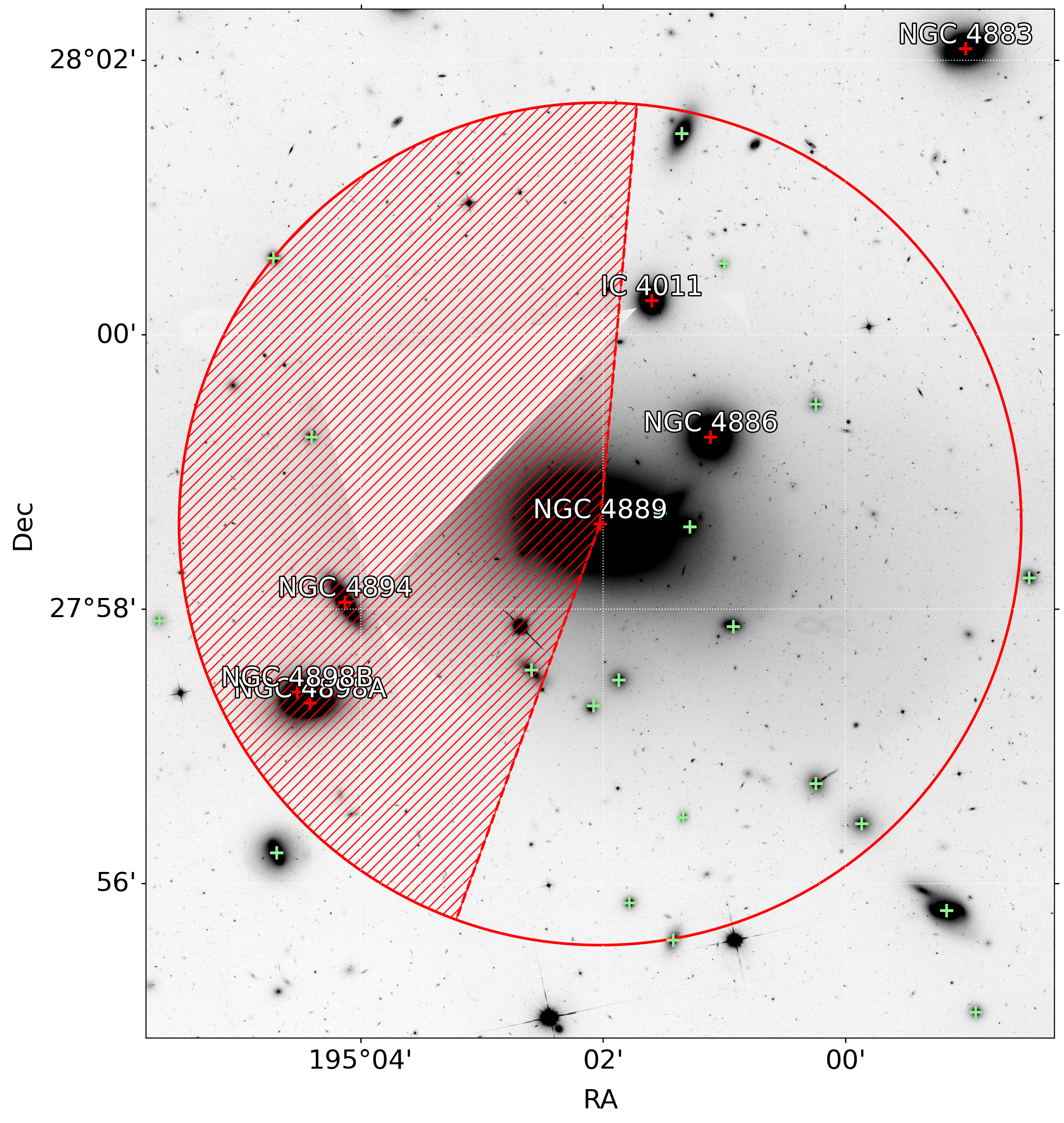}
    \caption{HST F814W image of NGC~4889 region.}
    \label{fig:Coma_NGC 4889_hst}
\end{subfigure}
\vspace{4 pt}
\begin{subfigure}[t]{0.49\linewidth}
    \centering
    \includegraphics[width=1\linewidth]{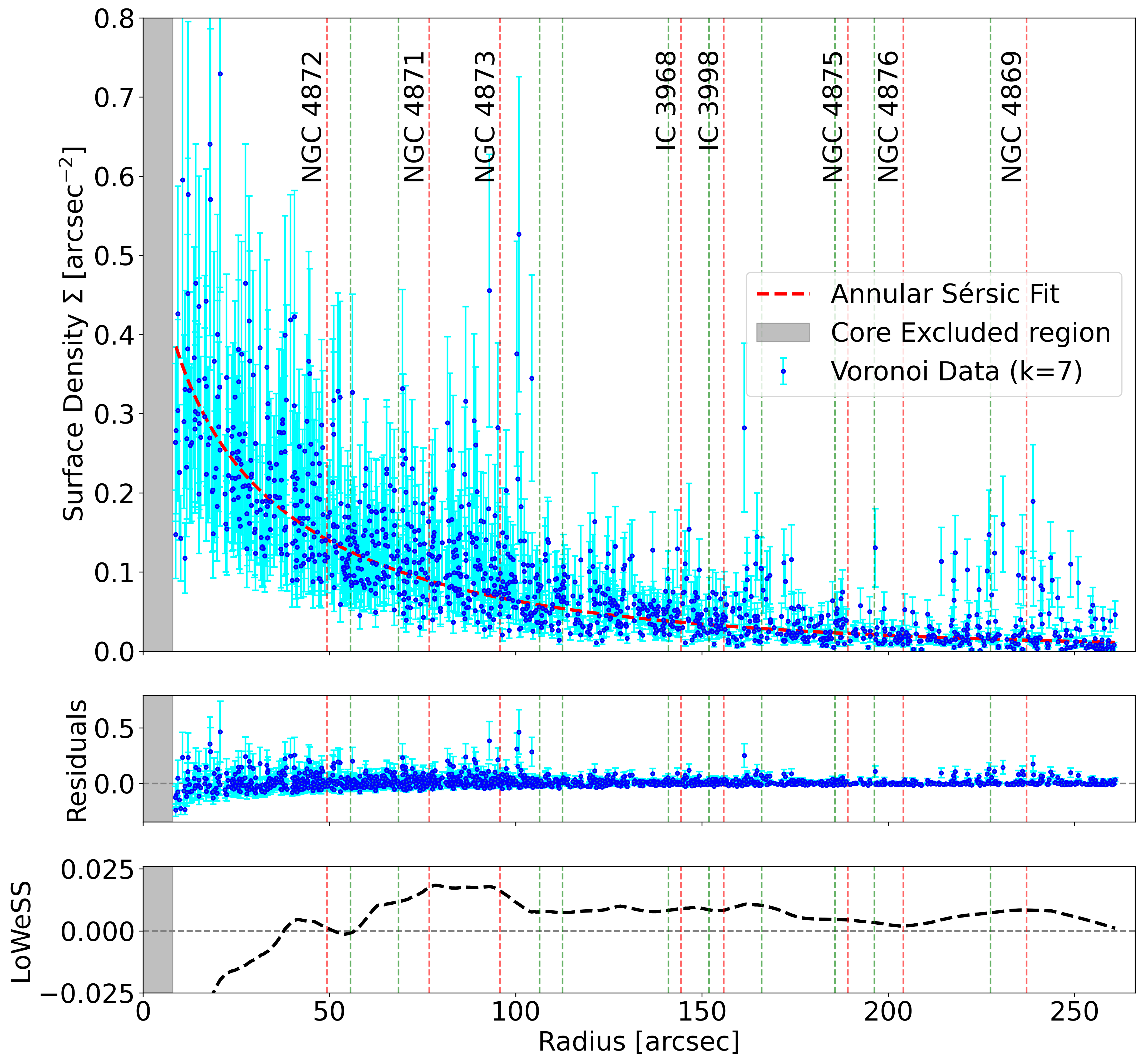}
    \caption{Radial density profile of GCs out to $9R_e$ for NGC~4874.}
    \label{fig:Coma_NGC 4874_Voronoi}
\end{subfigure}
\hspace{2 pt}
\begin{subfigure}[t]{0.49\linewidth}
    \centering
    \includegraphics[width=1\linewidth]{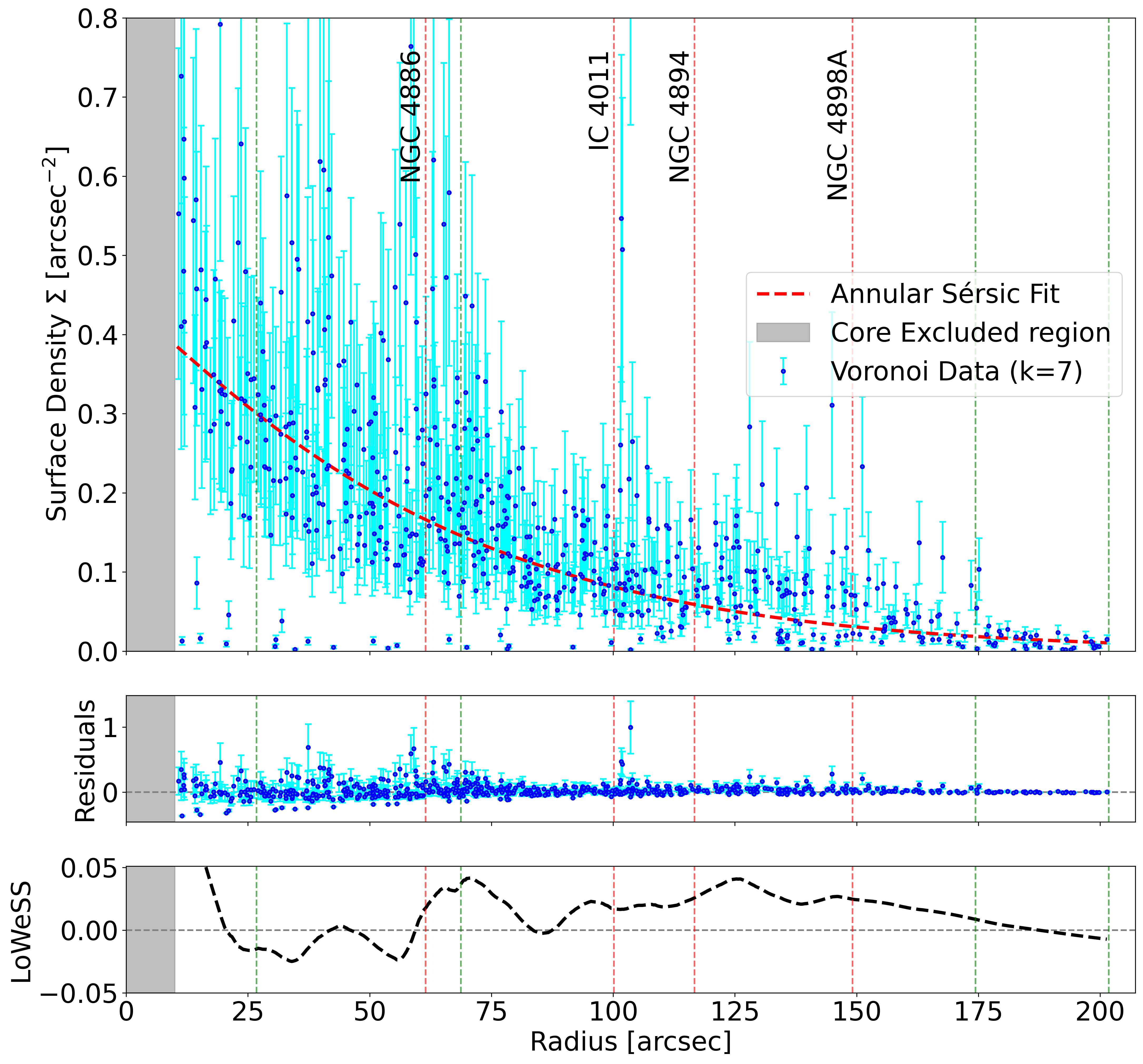}
    \caption{Radial density profile of GCs out to $9R_e$ for NGC~4889. }
    \label{fig:Coma_NGC 4889_Voronoi}
\end{subfigure}
\caption{(Upper panels) Combined HST F814W calibrated images of Coma BCGs. Images show location of brighter galaxies in proximity to the giant ellipticals. The red circle shows approximate extend of galaxy ($8R_e$) and red hatched region indicates area excluded from radial density analysis (due to HST pointing gap). Brighter galaxies are marked with red crosses and labeled, with green crosses indicating the remaining galaxies in our sample. (Lower panels) Radial density profile using Voronoi tessellation (see text) with `$k$' GC members per cell to reduce stochastic noise. Red and green Vertical dashed lines, indicate projected radial separation of sample galaxies (red: major (NGC/IC), green: minor, filtered to $<\!-18$ mag to reduce visual crowding). Grey shading $<\!\ang{;;10}$ indicates the core data masking region. S\'{e}rsic fit to annular radial profile of GCs included for comparison, from which residuals are calculated. Locally weighted scatterplot smoothing (LoWeSS) of residuals show density peaks at radii corresponding to many of the bright galaxies.} 
\label{1D_Voronoi_radial_profile}
\end{figure*}
\section{Voronoi Radial Profile}\label{sec:1D_voronoi}
We first explore the rich GCS around the two BCGs in Coma. Both NGC~4874 and NGC~4889 have several large neighboring galaxies within $\lesssim~8\Reff$ of their centers, which we take to represent the approximate extent of the BCGs. As the purpose of this analysis is to examine the profile of the GCSs around the BCGs, we note that, in general, the surface density of GCs consists of three components: the GCS of the BCGs, a background intergalactic component, and the GCSs of the nearby galaxies, \eg as discussed recently by \citet{D'Abrusco_2025} in their analysis of GC populations in the Fornax cluster. By analyzing the radial profile of GCS about the BCGs we look to find evidence of imprints of the nearby galaxies in the BCG profile. 

We proceed to produce a radial density profile using Voronoi tessellation in a manner similar to that outlined by \citet{Dornan_2024}. The advantage of the Voronoi method is that no binning is required, as the resolution adapts to local object density. The method is therefore good for the asymmetric, clumpy, or non-circular distributions which we are exploring as it preserves substructure that might otherwise be washed out in annular averaging.

Our GC dataset includes the J2000 coordinates of each of our candidate GC. Using these coordinates for each GC we assign a Voronoi cell, defining a region of space closer to that point than to any other. We then approximate the surface density, $\rho$, for each cell as 
\begin{align}
    \rho_i = \frac{1}{A_i} 
\end{align}
where $A_i$ is the area of the Voronoi cell around the object.

To implement this, we use the \texttt{Voronoi} method from the Python package \texttt{scipy.spatial}. After validating each cell and masking the radial area around the galaxy of interest, the densities are calculated. As noted by \citet{Dornan_2024}, radial profiles derived from individual Voronoi cells ($k=1$) can be highly noisy. Given the large number of cells available here, we therefore construct higher-order Voronoi tessellations by merging neighboring cells into groups of odd membership, $k$, which provides a controllable smoothing scale while preserving the adaptive nature of the method. In practical terms, we consider $k=\{3,5,7,9\}$, which provide progressively smoother density estimates, and adopt $k=7$ for the fiducial radial-profile analysis as a compromise between spatial resolution and noise suppression. The merging of $k$ cells is achieved using the unsupervised learner \texttt{NearestNeighbors} (part of the \texttt{sklearn.neighbors} package), ordering the groups by density. The surface density of each merged Voronoi cell is then  
\begin{align}
    \rho_j &= \frac{k}{\sum_{n=1}^k A_n}
    \intertext{with Poisson uncertainty}
    \sigma_j &= \frac{\rho_j}{\sqrt{k}}.
\end{align}
As a validation check, we compared the Voronoi-derived radial profile with the corresponding annular profile and found consistent large-scale behavior. Residuals were then defined relative to the annular S\'{e}rsic fit in order to highlight localized departures from the smooth radial trend. As a further robustness check, we also tested S\'{e}rsic + constant and S\'{e}rsic + linear forms for the annular radial profile, motivated by the possibility that the outskirts may include an additional large-scale background component, such as intracluster GCs. The constant term was not favored, converging to a negligible value, whereas a linear background term improved the formal fit in some cases. However, the additional linear term was partly degenerate with the S\'{e}rsic component, particularly for NGC~4889, so we do not interpret it as a unique physical decomposition. Importantly, these alternative parameterizations do not qualitatively change the main residual features or the broader trends discussed below. We therefore retain the single-S\'{e}rsic fit as an empirical description of the radial trend.

We present the Voronoi radial surface density around the dual Coma BCGs NGC~4874 and NGC~4889 in \cref{1D_Voronoi_radial_profile}. The top panels show HST cutouts, centered on NGC~4874 (\cref{fig:Coma_NGC 4874_hst}) and NGC~4889 (\cref{fig:Coma_NGC 4889_hst}), with other galaxies marked for reference. Brighter galaxies are marked with red crosses and labeled, with green crosses indicating the remaining galaxies in our sample, defined in Section \ref{ssec:gal_sample}. The red outlined circle indicates the galaxy $8\Reff$ used for the BCG. The hatched segment shown overlaid on NGC~4889 indicates the area excluded from the radial profile analysis due to the gap in HST data.

The lower panels of \cref{1D_Voronoi_radial_profile} show the corresponding Voronoi radial density profiles for NGC~4874 and NGC~4889, computed with the fiducial choice $k=7$, corresponding to 7 GCs per merged cell. The projected radial distances of the brighter galaxies (filtered to \qty{<18}{\mag} to reduce clutter) are indicated by dashed vertical lines, together with the annular S\'{e}rsic fit to the GC radial surface-density distribution. Residuals relative to this fit are shown beneath the main profile. To highlight coherent structure in these residuals, we also show a locally weighted scatterplot smoothing model \citep[LoWeSS;][]{Cleveland_1979}. Residuals relative to the annular S\'{e}rsic fit are shown below the main profile. To aid visualization of coherent structure in these residuals, we also plot a locally weighted scatterplot smoothing model \citep[LoWeSS;][]{Cleveland_1979}. This was implemented using the Python \texttt{statsmodels} package, which at each radius fits a local weighted linear regression to the nearest fraction of points in projected radius. We adopt a smoothing fraction of 0.1, such that each local fit uses the nearest 10\% of the residual points, weighted by their distance in radius from the evaluation point, with three robustness iterations to reduce the influence of outliers. The LoWeSS curve is shown only as a visual guide and is not used in the quantitative analysis.

Analysis of the surface-density profiles in \cref{1D_Voronoi_radial_profile} reveals localized departures from the smooth background trend in the vicinity of several bright galaxies. In some cases, excesses in the residual profiles are spatially offset from the projected galaxy centers. For example, in the vicinity of NGC~4889 (\cref{fig:Coma_NGC 4889_Voronoi}), the nearby elliptical NGC~4886 lies $\sim\!10$~arcsec from the nearest residual peak. This offset may reflect environmental substructure, projection effects, or contamination from the extended GCS of neighboring massive galaxies, rather than uniquely indicating environmental interactions associated with NGC~4886 itself.

\begin{deluxetable}{lccccccccc}
\tablecaption{Galaxy GC specific frequencies: Detections\label{tab:raw_GCSF_data_det}}
\tablehead{
    \colhead{Galaxy} & \colhead{MV (mag)} & \colhead{$\psub{N}{corr}$} & \colhead{$\psub{S}{N,obs} (\pm)$} & \colhead{$\psub{S}{N,pred}$} & \colhead{$\delta_f (\pm)$} & \colhead{$Z$} & \colhead{edge\_flag} & \colhead{nearest BCG} & \colhead{BCG dist. (")} 
}
\startdata
NGC 4889 & -23.10 & 4790(82) & 2.75(0.05) & 1.90 & 0.45(0.02) & 58.7 &  & NGC 4889 & 0.0 \\
NGC 4874 & -22.60 & 6176(90) & 5.64(0.08) & 1.41 & 3.00(0.06) & 68.6 &  & NGC 4874 & 0.0 \\
IC 4051 & -21.66 & 1666(43) & 3.61(0.09) & 0.79 & 3.54(0.12) & 38.5 &  & NGC 4889 & 623.9 \\
NGC 4869 & -21.22 & 223(21) & 0.72(0.07) & 0.62 & 0.17(0.11) & 10.6 &  & NGC 4874 & 237.1 \\
NGC 4908 & -21.11 & 44(14) & 0.16(0.05) & 0.58 & -0.73(0.08) & 3.2 &  & NGC 4889 & 622.3 \\
IC 4045 & -20.83 & 107(11) & 0.50(0.05) & 0.51 & -0.03(0.10) & 9.3 &  & NGC 4889 & 675.2 \\
IC 4042 & -20.80 & 88(11) & 0.42(0.05) & 0.51 & -0.17(0.11) & 7.9 &  & NGC 4889 & 459.7 \\
NGC 4871 & -20.75 & 146(29) & 0.73(0.14) & 0.50 & 0.45(0.29) & 5.0 &  & NGC 4874 & 76.8 \\
NGC 4906 & -20.73 & 94(11) & 0.48(0.05) & 0.50 & -0.03(0.11) & 8.8 &  & NGC 4889 & 461.0 \\
NGC 4873 & -20.67 & 109(28) & 0.59(0.15) & 0.49 & 0.21(0.31) & 3.9 &  & NGC 4874 & 95.8 \\
IC 3973 & -20.62 & 42(7) & 0.24(0.04) & 0.48 & -0.50(0.09) & 5.7 &  & NGC 4874 & 277.8 \\
NGC 4867 & -20.37 & 58(9) & 0.41(0.07) & 0.45 & -0.08(0.15) & 6.3 & edge & NGC 4874 & 274.2 \\
LEDA 44616 & -20.29 & 32(8) & 0.24(0.06) & 0.44 & -0.46(0.14) & 3.8 &  & NGC 4874 & 319.5 \\
IC 3976 & -20.28 & 38(7) & 0.30(0.05) & 0.44 & -0.33(0.11) & 5.8 & edge & NGC 4874 & 401.6 \\
NGC 4875 & -20.13 & 43(9) & 0.38(0.08) & 0.44 & -0.12(0.18) & 4.8 &  & NGC 4874 & 189.2 \\
LEDA 44792 & -20.11 & 57(12) & 0.51(0.11) & 0.43 & 0.18(0.25) & 4.7 &  & NGC 4889 & 428.1 \\
LEDA 44809 & -20.05 & 73(10) & 0.70(0.09) & 0.43 & 0.61(0.22) & 7.3 &  & NGC 4889 & 463.0 \\
LEDA 44723 & -19.79 & 20(6) & 0.24(0.07) & 0.44 & -0.45(0.17) & 3.2 &  & NGC 4889 & 360.1 \\
LEDA 44656 & -19.77 & 62(13) & 0.77(0.16) & 0.44 & 0.75(0.37) & 4.7 &  & NGC 4874 & 166.0
\enddata
\tablecomments{Systems passing the detection criterion $Z\geq3$; quoted uncertainties are $1\sigma$.}
\end{deluxetable}

Similarly, departures below the LoWeSS trend are observed near several systems. In the NGC~4874 field (\cref{fig:Coma_NGC 4874_Voronoi}), the ellipticals NGC~4872 and NGC~4876 coincide with regions of reduced residual surface density. Given the dominance of the BCG halo in this region, such deficits may arise from the overlap of multiple GC populations and spatial variations in detection completeness, and we therefore interpret this with caution.

Overall, while some galaxies exhibit localized enhancements in GC surface density that rise above the smooth BCG and intracluster background (e.g., NGC~4873), others appear strongly diluted by the surrounding cluster population (e.g., NGC~4876). These variations highlight the complex, overlapping nature of GCS in the Coma core.

The presence of spatial offsets and localized deficits is suggestive of asymmetries in the projected GC distribution. While such features are qualitatively consistent with environmental processing and tidal interactions in the cluster core, they are not uniquely diagnostic. We therefore defer detailed physical interpretation and examine the two-dimensional distribution of GCs in Section~\ref{sec:2D_voronoi_density}, where projection effects and substructure can be assessed more directly. Before doing so, we first investigate the GC specific frequency for galaxies in the Coma cluster to explore the validity of this empirical parameter.

\section{GC Specific Frequency of Coma Cluster Galaxies}\label{sec:Coma_gal_GCSF}

The relationship between the luminosity of an elliptical galaxy, which are prevalent in the Coma cluster, and its satellite GC population, as detailed by \citet{Harris_1981} is known as the `specific frequency' \SN. Similar specific frequency relationships also exist for spirals and irregular galaxies \citep[see][]{Harris_2013}. The GC specific frequency gives an estimate of the number of GCs per unit of galactic luminosity. For galaxies where the absolute luminosity is known we can therefore use this to estimate the expected number of GCs and compare this to the observed number. However, for the reasons outlined above, the GC specific frequency may not be valid in a galaxy cluster environment, or at best will exhibit a greater scatter compared to galaxies in an isolated environment.

\subsection{Estimation of observed GC specific frequency, \texorpdfstring{$\SN$}{SN}}

We first compute a background-corrected count of GCs around each galaxy \psub{N}{GC,corr} from which we determine \SN.

The specific frequency \SN parametrizes the well established empirical relationship between galaxy luminosity (\MV) and number of GCs (\psub{N}{GC}) of elliptical galaxies \citep{Harris_1981}. We employ this to first explore the GCS population of the brighter Coma galaxies within our observed region, \ie those with $\MV\lesssim-19.6$ mag. This magnitude cut ensures inclusion of all 35 NGC/IC catalog galaxies within the surveyed region, together with 12 additional galaxies from 2MASX, LEDA, and SDSS.

From our observational catalog of GCs built with HST data, we determine the count of GCs within a projected radius of $8\Reff$ of each of our target galaxies. For consistency, we used the homogeneous values of effective radius (\Reff) for each galaxy derived from SDSS photometry, as outlined in Section~\ref{sec:p2_methods}. The observed GC counts within $\psub{R}{in}=8\Reff$ of each galaxy were corrected for local background contamination by measuring GC counts in an outer annulus ($12\Reff < \psub{R}{ann} < 20\Reff$) and subtracting the area-scaled contribution from the inner counts. We tested the sensitivity to aperture choice and found little dependence on this selection (see Appendix~\ref{app:GCSF-robustness}). Poisson uncertainties on the observed GC sample count, including propagation through the background subtraction, were also calculated. Given the high projected galaxy density in Coma, no attempt was made at this stage to assign individual GCs to specific host galaxies. We flag coverage on galaxies close to the outer and inner edges of HST pointings for later diagnostics. At this stage, we exclude galaxies outside the HST pointing footprint for which \SN would likely be unreliable, even though they may imprint on the GC surface density.

With our background-corrected count of GCs for each galaxy, we compute the corresponding \SN\ via
\begin{equation}\label{eq:sf}
        \SN=\psub{N}{GC,corr}10^{0.4(\MV+15)}.
\end{equation}

\begin{figure}[t!]
    \centering
    \includegraphics[width=0.99\linewidth]{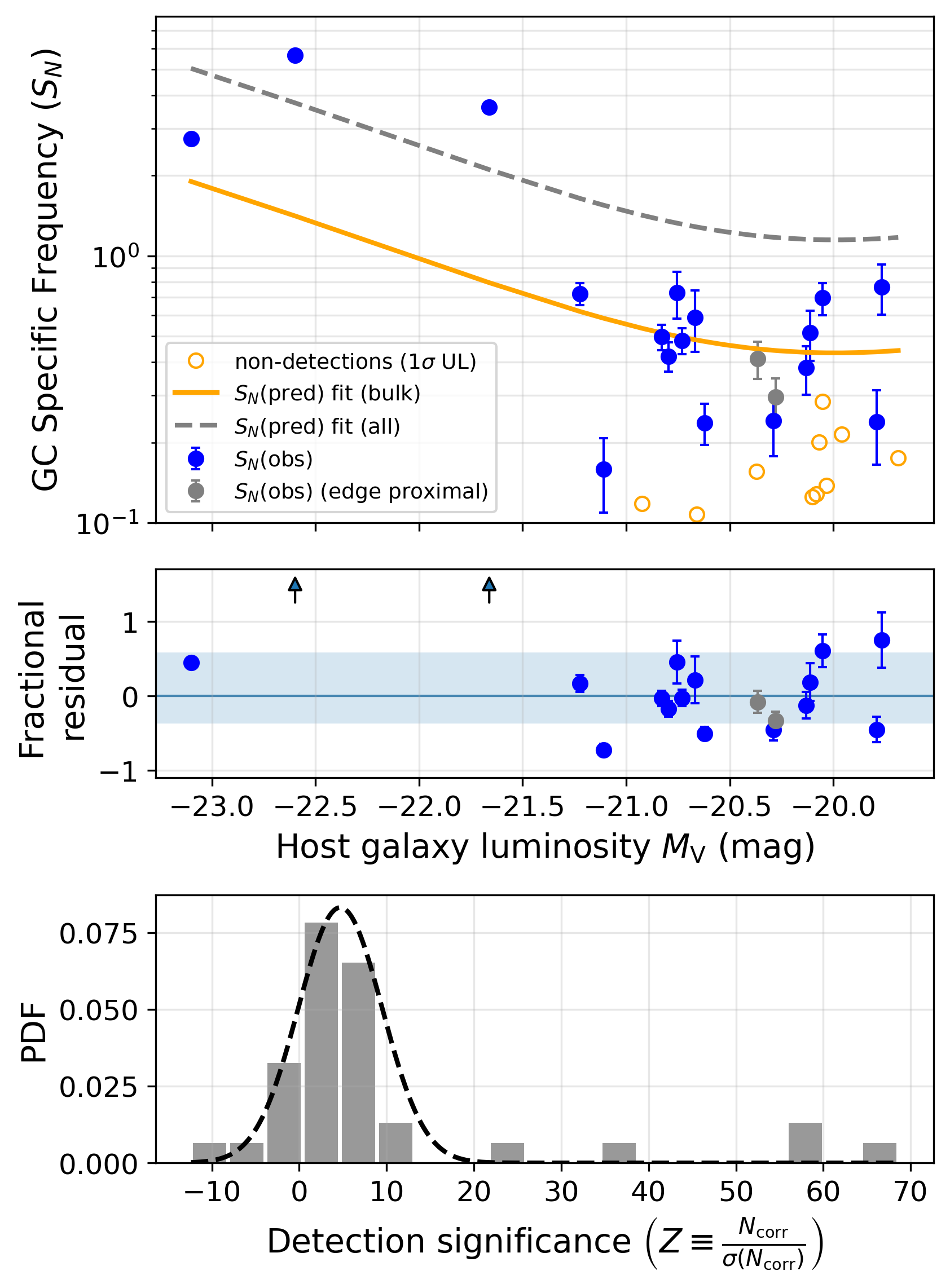}
    \caption[Specific frequency of globular clusters as a function of host galaxy luminosity.]{Specific frequency of globular clusters as a function of host galaxy luminosity. (Top) Observed GC specific frequency $\SN(\rm obs)$ versus \MV for Coma cluster galaxies (blue points), with edge-proximal systems shown in gray and non-detections indicated as open symbols. The curves show the normalized baseline relation derived from the Virgo cluster sample \citep{Peng_2008}. The dashed curve shows the normalization obtained when GC-rich systems (BCGs and IC 4051) are included, while the solid curve shows the `bulk' normalization anchored to the mid-luminosity population $(-21.5\gtrsim\MV\gtrsim\qty{-19.6}{\mag})$. We adopt the bulk normalization in subsequent analysis to avoid bias from known GC-rich outliers. (Middle) Fractional residuals $(\psub{S}{N,obs}-\psub{S}{N,pred})/\psub{S}{N,pred}$ relative to the baseline relation. The shaded region indicates the $\pm1\sigma$ intrinsic scatter of the relation ($\sigma\psub{\log{S}}{N}=0.2$ dex). Error bars reflect measurement uncertainties only. Galaxies lying systematically below the expected relation are examined further in subsequent sections. (Bottom) Distribution of the detection significance $Z\equiv\psub{N}{corr}/\sigma(\psub{N}{corr})$, with a Gaussian fit to the central core shown for reference. The extended positive and negative tails reflect systems with unusually rich or deficient GC populations, respectively.}
    \label{fig:Coma_SF_v2}
\end{figure}

We then use the GC specific frequency, \SN, in bins of \MV\ as determined for the Virgo cluster \citep[see][table 3]{Peng_2008}, from which we derive a third-order polynomial that reproduces the published relation to within the typically quoted scatter \citep[\eg][]{Huang_2021}. We determine the best-fit vertical normalization of this relation for our Coma galaxy data using a weighted log-space fit. We present the specific frequency for our Coma galaxies in \cref{fig:Coma_SF_v2}. The top panel shows GC specific frequency as a function of galaxy absolute $V$-band magnitude. Blue markers indicate detections where \SN\ is positive and the significance
\begin{equation}\label{eq:sn_signif}
Z \equiv \frac{\psub{N}{GC,corr}}{\sigma(\psub{N}{GC,corr})},
\end{equation}
exceeds 3. Galaxies in proximity to the pointing edges, using the same conservative quality threshold, are indicated in gray. Upper limits for non-detections outside this threshold are shown as open orange symbols. The curves show the Virgo cluster \SN-$\MV$ relation from \citet{Peng_2008}, vertically renormalized to match our Coma data. The dashed curve corresponds to a normalization fitted to all detections, while the solid orange curve shows the normalization obtained when excluding the BCGs and IC~4051, highlighting the influence of these GC-rich systems on the overall scaling. Our $\SN$ estimates for the BCGs are broadly consistent with previous work \citep[\eg][]{Cho_2016}, which finds elevated $\SN$ values for cD galaxies in Coma. The galaxies passing our detection threshold are listed in Table~\ref{tab:raw_GCSF_data_det}. 

The middle panel of \cref{fig:Coma_SF_v2} shows the fractional residual of the background corrected \SN\ of our observed detections to the normalized Peng-\SN\ curve, \ie 
\begin{align}
    (\psub{S}{N,corr}-\psub{S}{N,pred})/\psub{S}{N,pred}
\end{align}.
The shaded band on this plot indicates the $\pm1\sigma$ intrinsic scatter in $\log\SN$($\sigma\approx0.19$ dex), with bootstrap confidence intervals derived from 5000 resamples. 

We tested the stability of the \SN residuals under variations in inner aperture, background annulus, and magnitude limit. The best-fit bulk normalization and the inferred intrinsic scatter remain stable ($\sigma \approx 0.2~{\rm dex}$) across all configurations, with high overlap between selections under bootstrap resampling. Restricting the analysis to magnitudes brighter than F814W~\qty{<25.8}{\mag} yields consistent results. We therefore conclude that the deficit signal is robust to reasonable methodological variations.

\begin{figure*}[ht]
    \centering
    \includegraphics[width=1\linewidth]{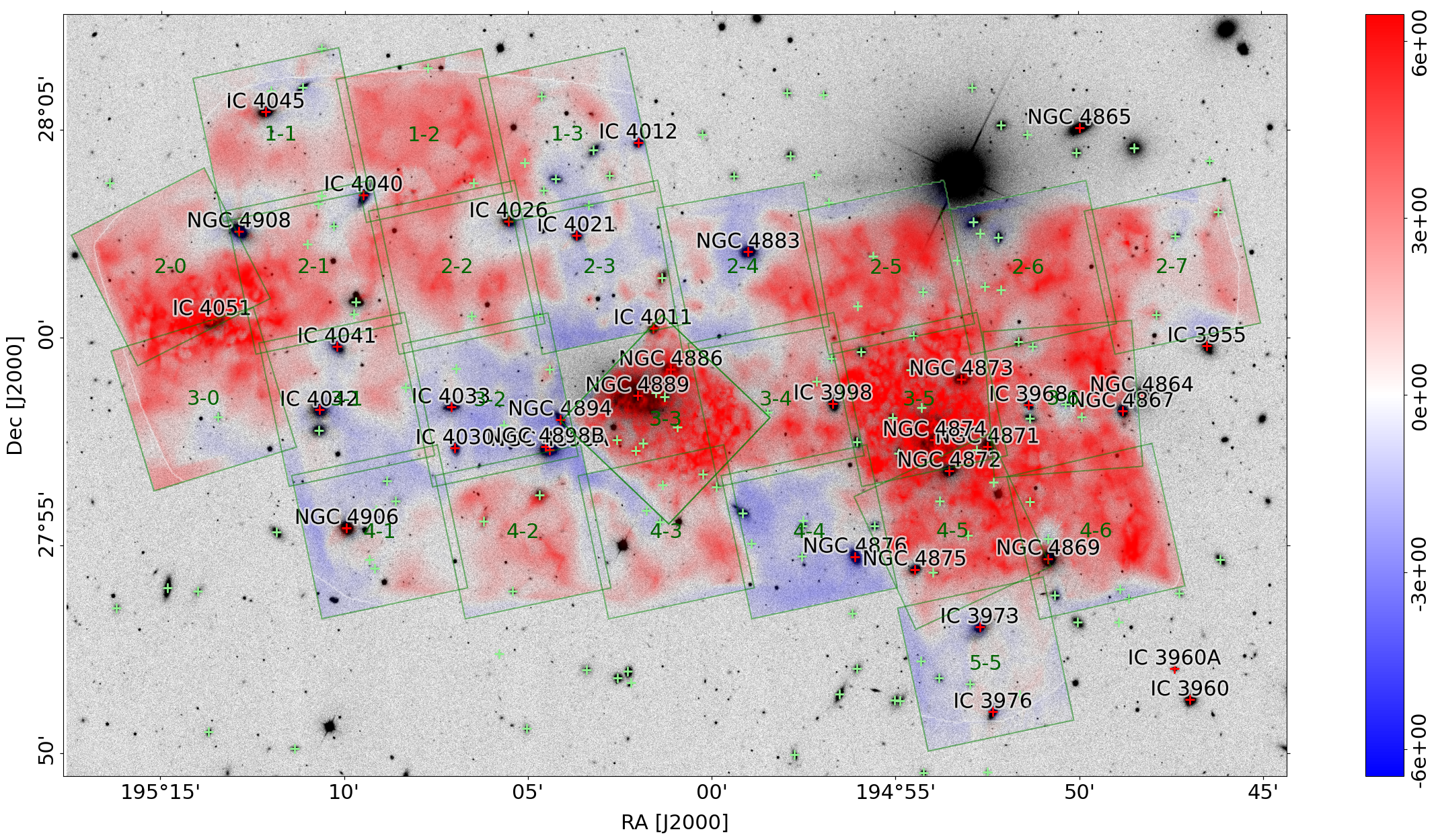}
    \label{fig:Coma_2D_Voronoi_Signif_Coma_cluster}
    \caption[Residual significance of Voronoi density map of the Coma cluster.]{Residual significance of Voronoi density map of the Coma cluster. 2D Voronoi density for the observed region of the Coma cluster with legends as for \cref{fig:Coma_NGC 4874_2D_Vor_density}. Result overlaid on SDSS $g$-band image, with HST pointing outlines shown for reference. Simulation parameters as described in the text. The thin white contour marks the boundary of the region with sufficient data support from the Voronoi tessellation and simulation realizations. Density estimates outside this region are masked. Over and under-densities of GCs are clear.}
    \label{fig:Coma_cluster_2D_Vor_density}
\end{figure*}

After excluding GC-rich outliers (the BCGs and IC 4051), the majority of Coma galaxies exhibit GC specific frequencies consistent with the Virgo baseline within the intrinsic scatter. Most galaxies in the bulk, non-proximal BCG sample follow this baseline relation, with the median fractional residual for clean detections consistent with zero (median $=-0.03$). In contrast, galaxies classified as non-detections exhibit systematically negative upper-limit residuals (median $=-0.73$), consistent with a distinct low-$\SN$ tail. Several of these galaxies are formally classified as upper limits, but their limits are consistent with severe GC depletion. This heterogeneous behavior suggests that environmental processing affects only a subset of galaxies rather than the population as a whole.

Importantly, we have confirmed this result is not driven by completeness or aperture choice (see Appendix~\ref{app:GCSF-robustness}). The normalization offset and deficit fraction remain stable under variations in background estimation, radial aperture, and magnitude limit. To explore the nature of these deficits further, we next examine the two-dimensional GC surface-density field using a Voronoi-based approach.

\begin{figure*}[t!]
    \begin{subfigure}{0.49\linewidth}
        \centering
        \includegraphics[width=1\linewidth]{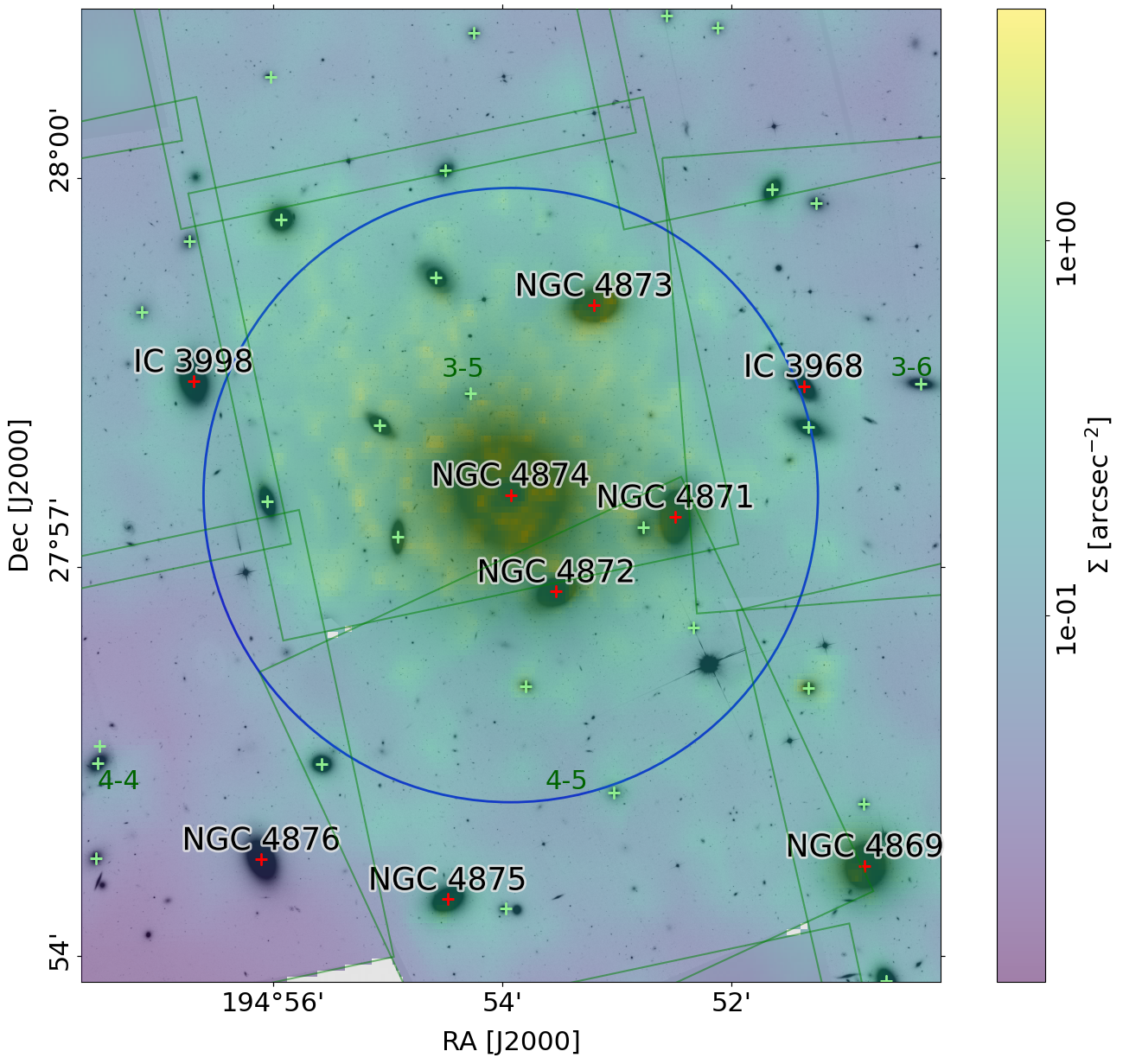}
        \caption{Observed}
        \label{fig:Coma_NGC 4874_2D_Vor_obs}
    \end{subfigure}
    \begin{subfigure}{0.49\linewidth}
        \centering
        \includegraphics[width=1\linewidth]{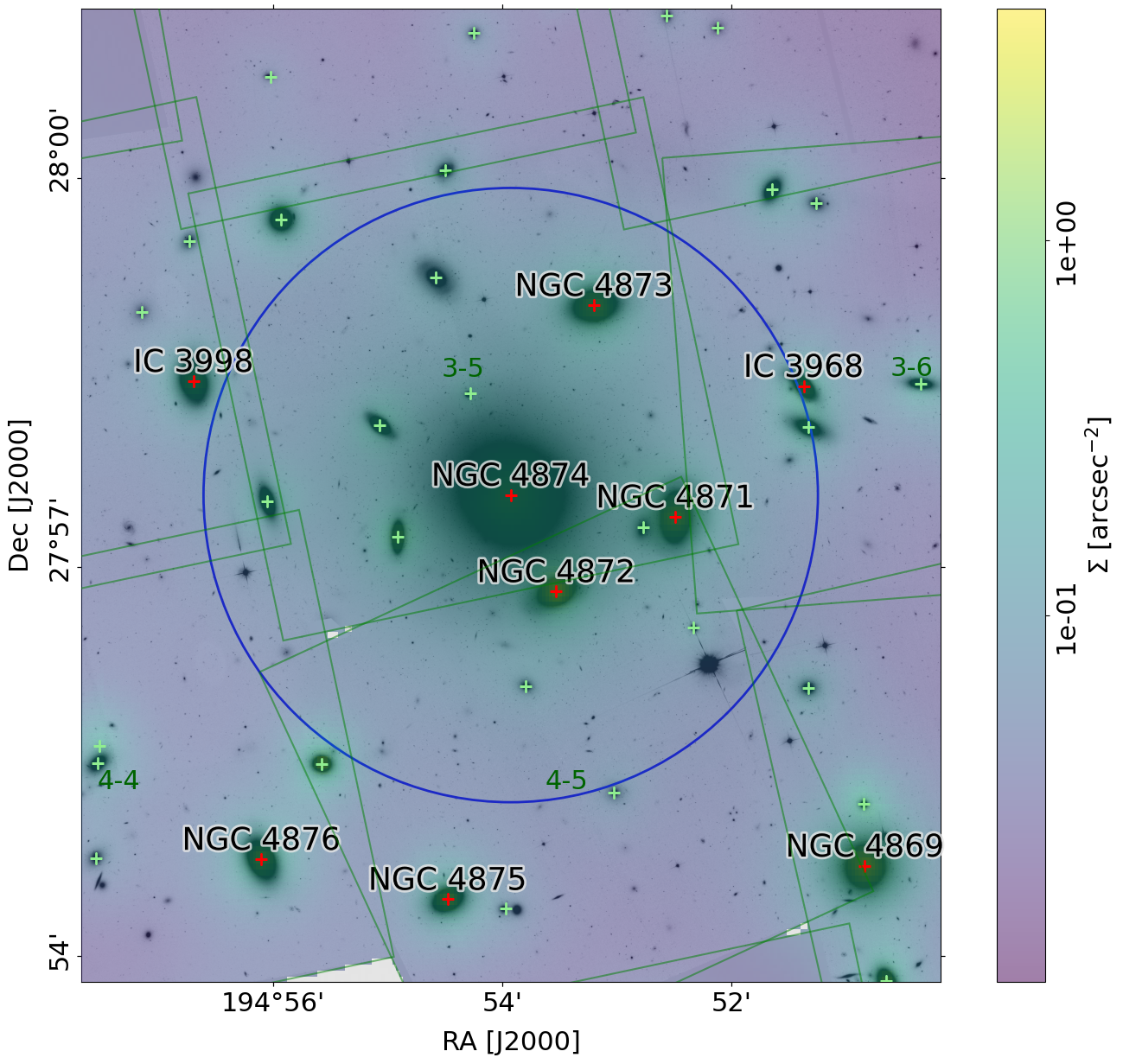}
        \caption{Simulated}
        \label{fig:Coma_NGC 4874_2D_Vor_sim}
    \end{subfigure}
    
    \vspace{5 pt}
    \begin{subfigure}{0.49\linewidth}
        \centering
        \includegraphics[width=1\linewidth]{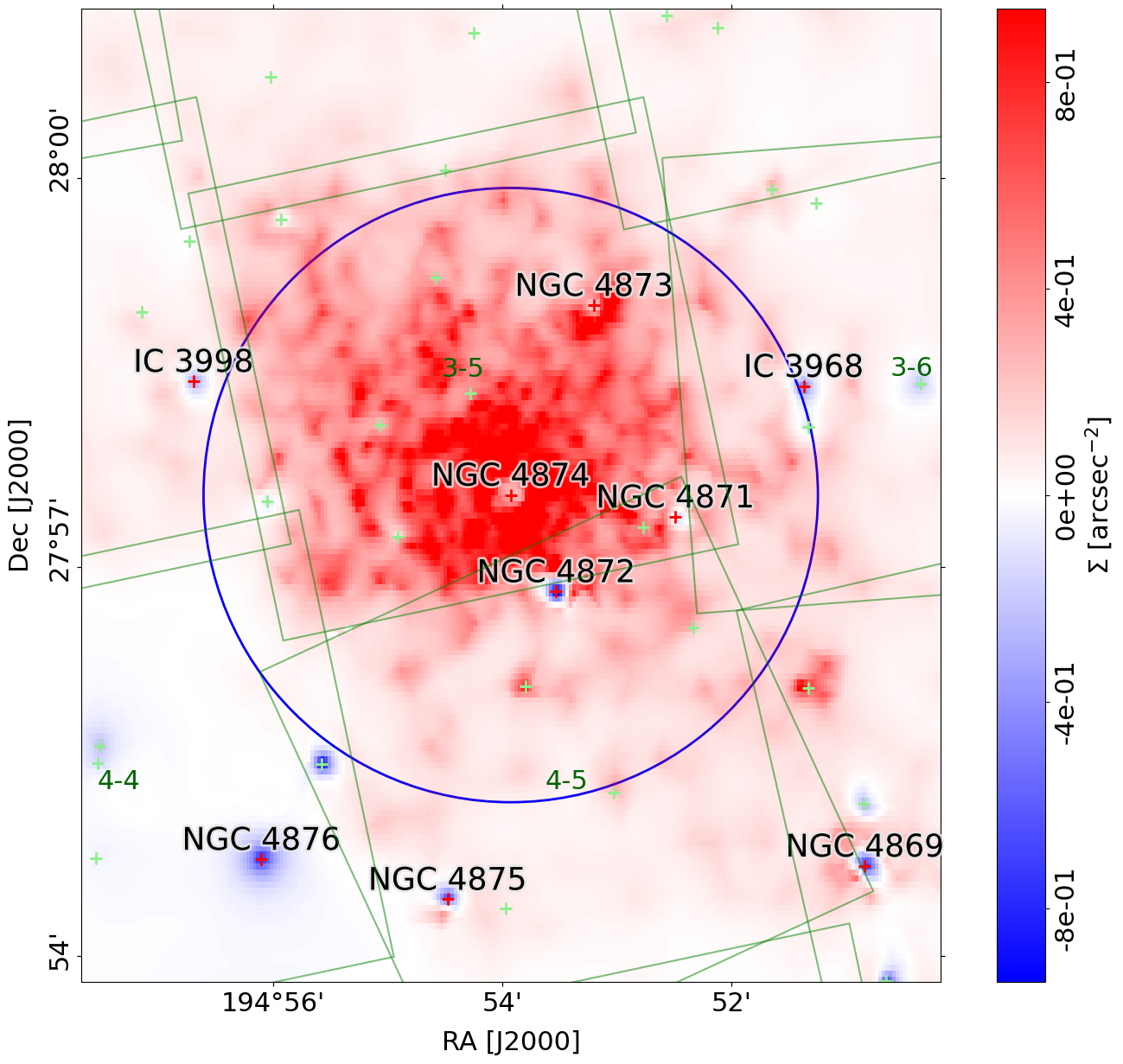}
        \caption{Residual}
        \label{fig:Coma_NGC 4874_2D_Vor_res}
    \end{subfigure}
    \hspace{5pt}
    \begin{subfigure}{0.47\linewidth}
        \centering
        \includegraphics[width=1\linewidth]{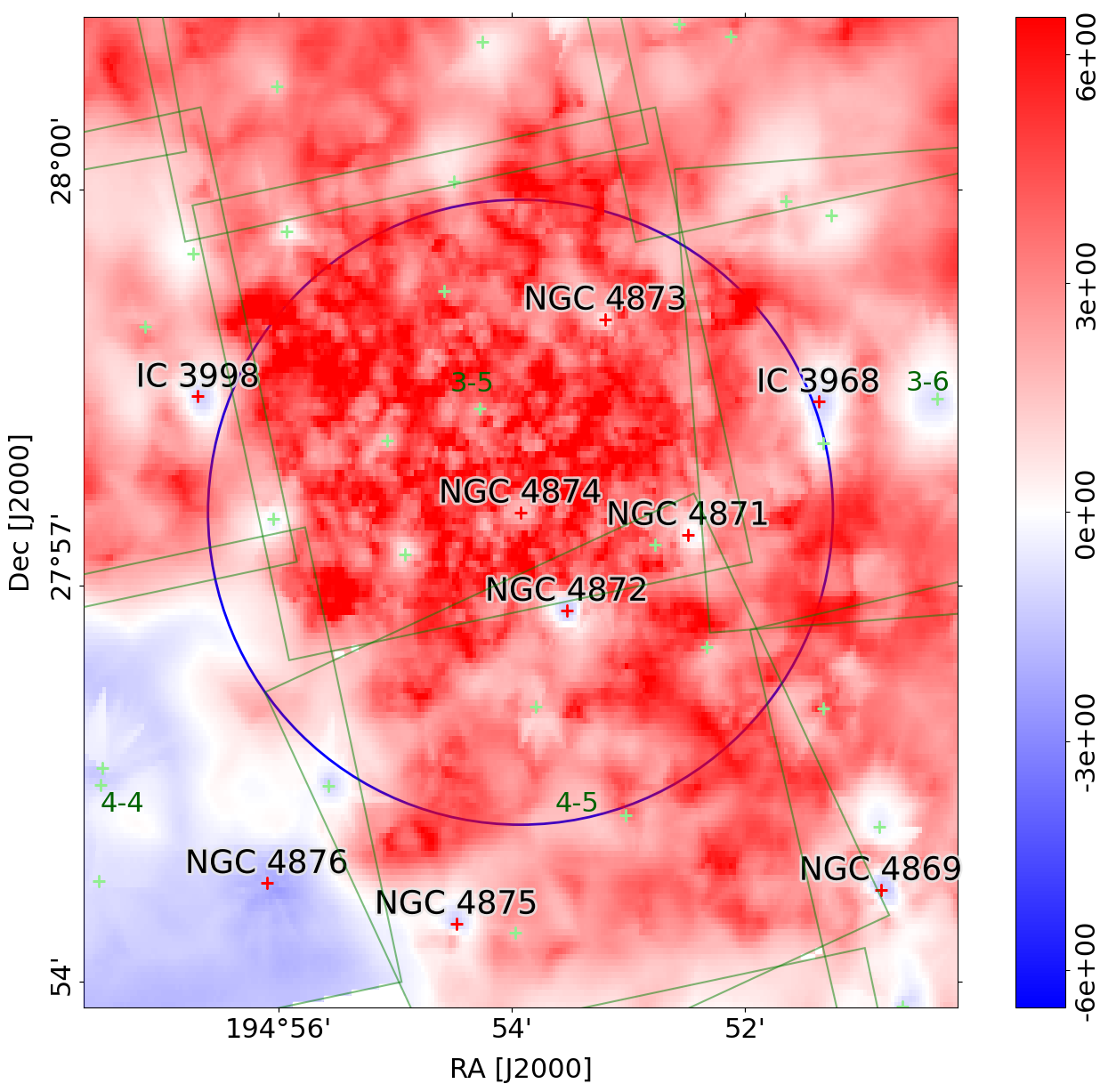}
        \caption{Significance}
        \label{fig:Coma_NGC 4874_2D_Vor_sig}
    \end{subfigure}
    \caption{2D Voronoi density in proximity to NGC~4874. Observed and simulated densities are overlaid on F814W filter HST / ACS images, with pointing outlines shown for reference. Simulation parameters as described in the text. Blue circle indicative of half-number effective radius of the GCS, \ReffGCS, for the central galaxy. Over and under-densities of GC are clear on both residual (c) and significance (d) frames.}
    \label{fig:Coma_NGC 4874_2D_Vor_density}
\end{figure*}

\section{2D Voronoi Density Grid}\label{sec:2D_voronoi_density} 

To investigate the spatial distribution of GC candidates across the full HST footprint, we extend the Voronoi approach described in Section~\ref{sec:1D_voronoi} to construct two-dimensional surface-density maps. These maps allow a direct comparison between the observed GC distribution and the expectations from galaxy-based GC system models, enabling the identification of localized excesses, deficits, and azimuthal asymmetries across the cluster field.

\subsection{Observed density map}
To create a 2D Voronoi density map from the observed data, we start with the coordinates of the GC candidates, and follow the same process outlined in Section \ref{sec:1D_voronoi} assigning candidates to tessellation cells with `$k$' members. The primary difference here was the full GC candidate dataset was included and HST pointing footprints were used for data masking. Interpolation to a fixed (RA, Dec) grid (of resolution $1400\times800$ within the area of interest) was achieved using \texttt{linear} interpolation for the density data, and \texttt{nearest} interpolation for the Poisson uncertainty.

To reduce sensitivity to the particular Voronoi tessellation produced by the discrete GC positions, we applied bootstrap resampling to the GC catalog. Specifically, $N_b=80$ bootstrap realizations of the GC candidate list were generated by randomly sampling the catalog with replacement while preserving the original sample size. A Voronoi density grid was constructed for each realization following the procedure above, and the resulting grids were aggregated to produce the final observed density map used for comparison with the simulations.

\subsection{Simulated density map}
To simulate the density of GCs within the observed region, we use the galaxy sample parameters extracted from archives as detailed in \ref{ssec:gal_sample}, including galaxies in an extended neighboring region outside the HST footprint. The parameters used for each simulated galaxy are the GCS S\'{e}rsic index, $n$, the GCS effective radius, $\ReffGCS$, in arcsec, and the host-galaxy absolute visual magnitude, $\MV$.

The GC specific frequency, $\SN$, inferred from $\MV$, sets the total GC population associated with each galaxy. The corresponding GCS S\'{e}rsic profile parameters ($\ReffGCS,n$) then define the spatial distribution of that population, where $\ReffGCS$ is the half-number radius and $n$ controls the profile shape. We use the inverse cumulative distribution function (CDF) of the S\'{e}rsic law to draw random projected radii for the simulated GCs, assigning each object a random azimuthal angle about the galaxy center.

For each galaxy, we generated multiple Monte Carlo realizations of this model population. In each realization, the underlying galaxy parameters and total GC population were held fixed, but the individual GC positions were redrawn stochastically from the S\'{e}rsic-based spatial distribution, producing a new discrete synthetic catalog.
\begin{figure*}[t!]
    \begin{subfigure}{0.49\linewidth}
        \centering
        \includegraphics[width=1\linewidth]{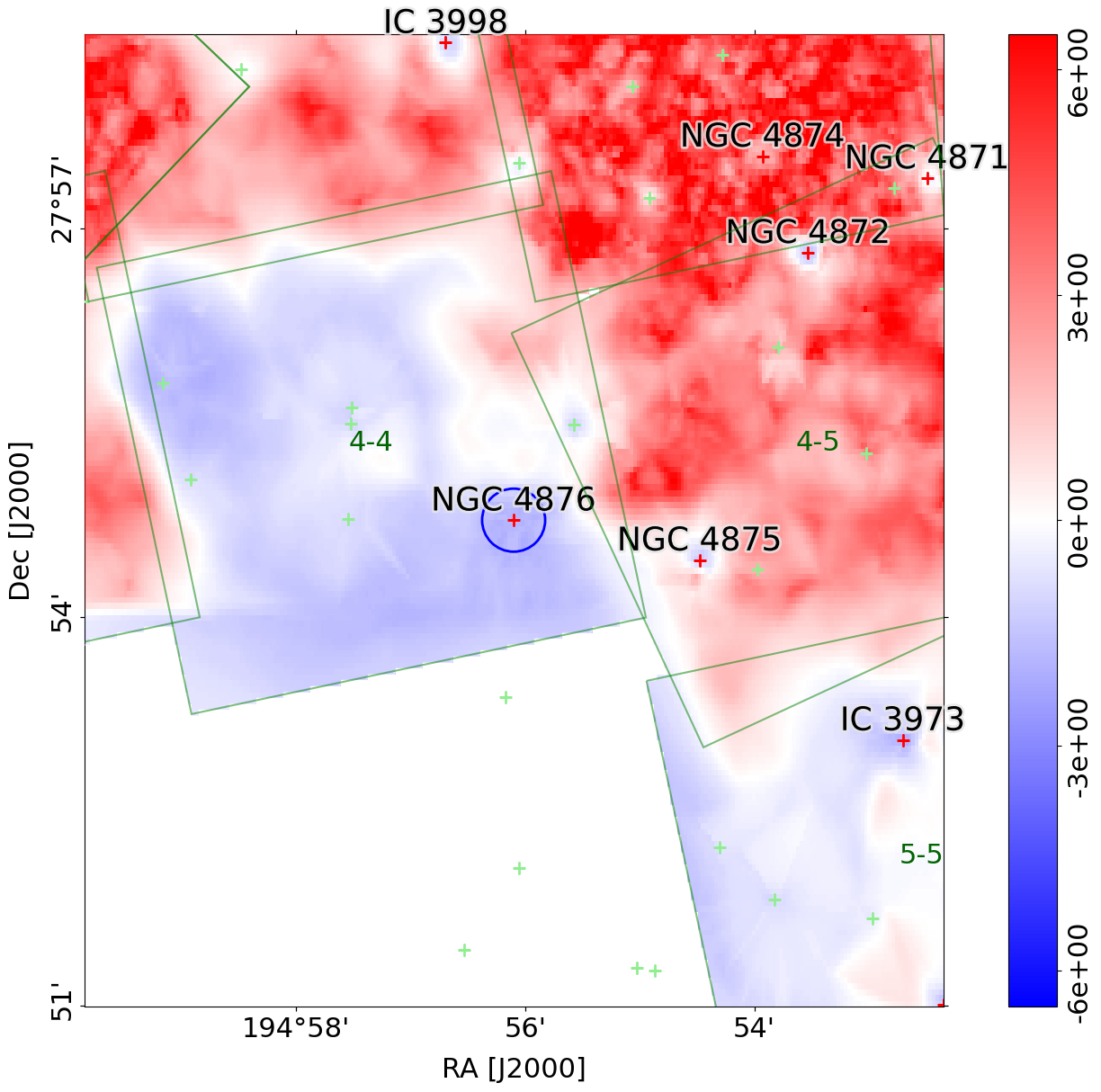}
        \caption{NGC 4876 residual significance}
        \label{fig:Coma_NGC 4876_2D_Vor_sig}
    \end{subfigure}
    \begin{subfigure}{0.49\linewidth}
        \centering
        \includegraphics[width=1\linewidth]{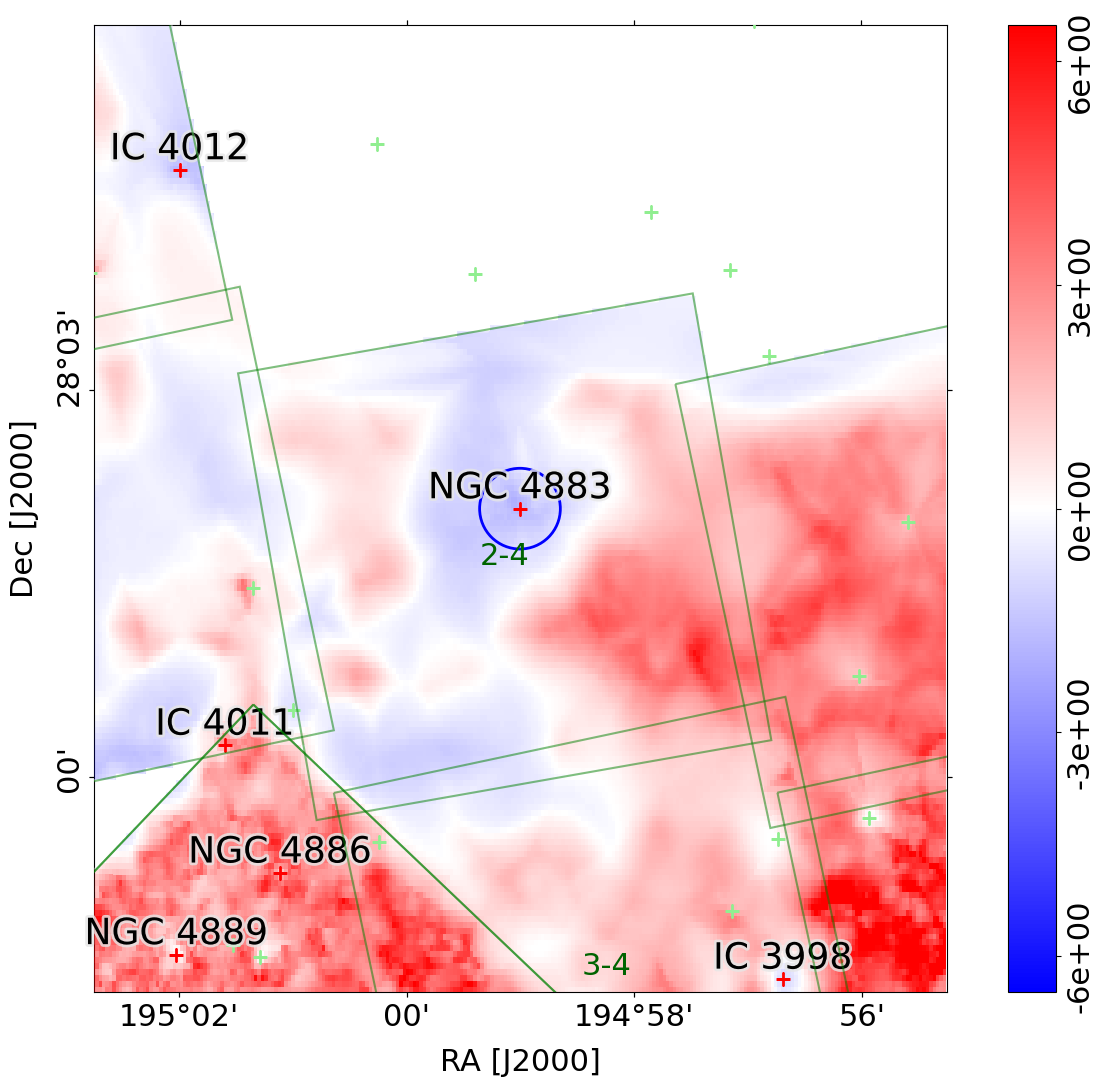}
        \caption{NGC 4883 residual significance}
        \label{fig:Coma_NGC 4883_2D_Vor_sig}
    \end{subfigure}
    \vspace{4 pt}
    \begin{subfigure}{0.49\linewidth}
        \centering
        \includegraphics[width=1\linewidth]{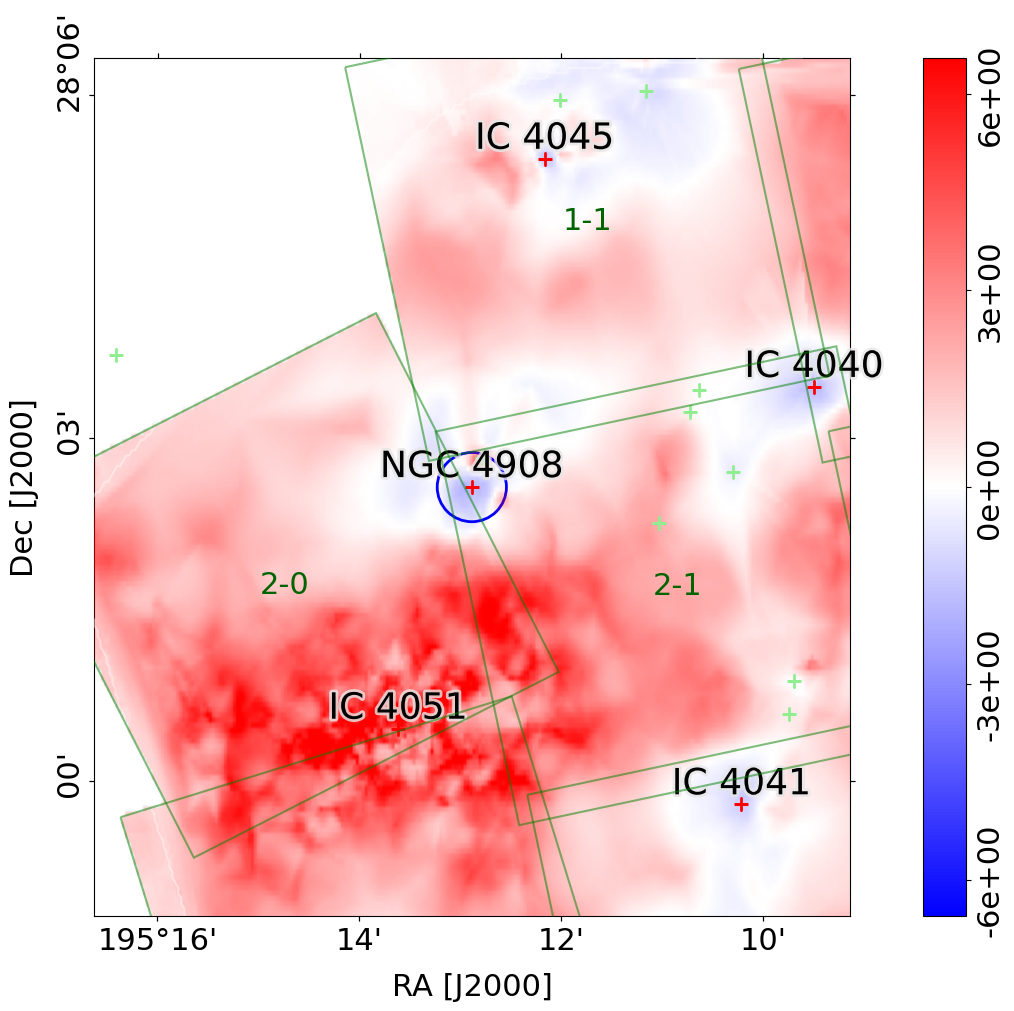}
        \caption{NGC 4908 residual significance}
        \label{fig:Coma_NGC 4908_2D_Vor_sig}
    \end{subfigure}
    \hspace{5 pt}
    \begin{subfigure}{0.48\linewidth}
        \centering
        \includegraphics[width=1\linewidth]{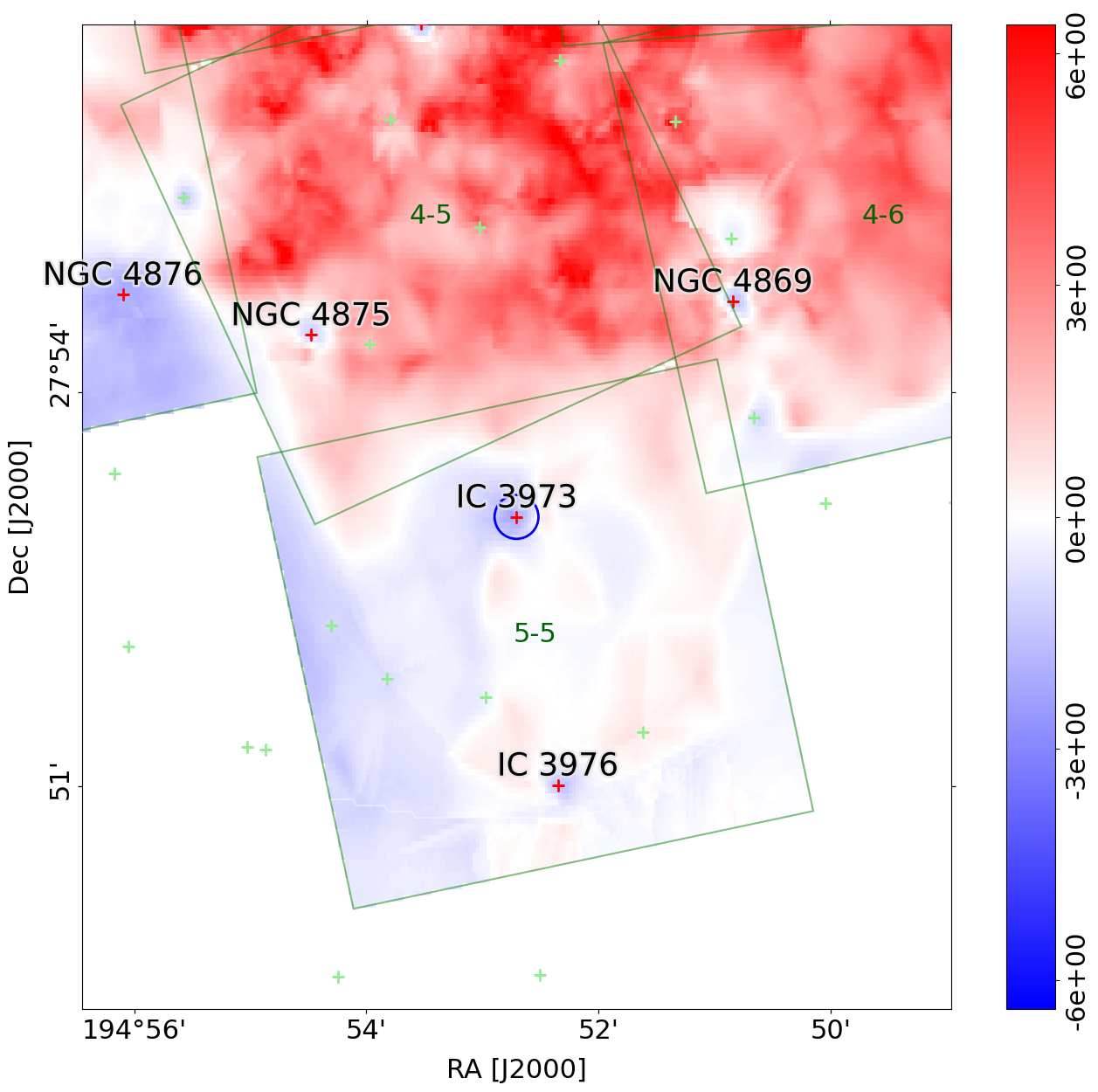}
        \caption{IC 3973 residual significance}
        \label{fig:Coma_IC 3973_2D_Vor_sig}
    \end{subfigure}
    \caption{2D Voronoi density residual significance, $S$, centered on NGC~4876, NGC~4883, NGC~4908 and IC~3973, identified in Section \ref{sec:Coma_gal_GCSF}. Legend and simulation parameters are consistent with those noted in \cref{fig:Coma_NGC 4874_2D_Vor_density} and discussed in the text.}
    \label{fig:Coma_other_gals_2D_Vor_density}
\end{figure*}

For GC-system modeling, we estimated the GC-system effective radius, $\ReffGCS$ as a scale multiple of the galaxy $\psub{R}{e,\star}$ using $\ReffGCS=3.0\,\psub{R}{e,\star}$ for the few non-ellipticals in the region, $3.2\,\psub{R}{e,\star}$ for ellipticals, and $4.8\,\psub{R}{e,\star}$ for large Es/BCG-like systems. These assumptions on GCS parameters are supported through previous surveys such as the SAGES Legacy Unifying Globulars and GalaxieS (SLUGGS), \citep{Kartha_2014}, and the Next Generation Virgo Cluster Survey (NGVS) \citep{Lim_2024} and studies of nearby galaxy groups identified by \citet{Hudson_2018} in the Canada-France-Hawaii Telescope Legacy Survey (CFHTLS) field. For the S\'{e}rsic index, $n$, we note that when analyzing the intracluster GC population in Coma, \citet{Peng_2011} varied the S\'{e}rsic index, $n$ from $1-4$ for fitting, but they use a S\'{e}rsic plus constant model fit to the radial profile across the whole cluster, centered on the BCG NGC~4874. They found a best fit S\'{e}rsic index of $n=1.3$. More recently, \citet{Lim_2025} fit S\'{e}rsic  laws across many early type galaxies finding typical indices $n$ spanning $~0.5-4$, with large objects tending to have shallower (lower-$n$) outer GCS profiles, consistent with BCGs having slightly flatter GCS radial profiles than ordinary ellipticals. From the fit of a S\'{e}rsic profile to the GCS of the BCGs shown in Section \ref{sec:1D_voronoi}, we obtained a best fit of $n\approx1.7\pm0.05$.  
 
Therefore, for simplicity, we use a S\'{e}rsic index of $\sim\!1.9$ throughout, with the exception of the BCGs, for which a S\'{e}rsic index of $\sim\!1.7$ was used.

In the production runs used for the final figures, we generated $n_r=200$ Monte Carlo realizations to obtain stable estimates of the mean density field and its variance. For each realization $i\in[1...n_r]$, we generated a simulated GC catalog using the S\'{e}rsic based sampler, computed its Voronoi density field, and interpolated this onto the fixed (RA,Dec) grid to get a 2D array $\rho^{(i)}(x,y)$. Interpolation to the grid was achieved using \texttt{linear} interpolation for the density data, and \texttt{nearest} interpolation for the Poisson uncertainty.

After all simulation runs, the $n_r$ density maps $\{\rho^{(i)}\}$ were stacked into a 3D array of shape $(n_r, N_y, N_x)$. From this stack, we computed the mean density and spatial standard deviation as
\begin{align}
    \langle \rho \rangle(x,y) &= \frac{1}{n_r}\sum_{i=1}^{n_r} \rho^{(i)}(x,y) \label{eq:2d_density}\\
    \sigma_\rho(x,y) &= \sqrt{\frac{1}{n_r-1}\sum_{i=1}^{n_r}\left[\rho^{(i)}(x,y)-\langle\rho\rangle(x,y)\right]^2} \label{eq:2d_sigma}
\end{align}
with the variance evaluated only in regions with sufficient ($>50\%$) simulation support across the realizations. A conservative floor was applied to the uncertainty grid to avoid spuriously small denominators when computing the residual significance.

\subsection{Comparison of observed and simulated densities}
As the GC candidate catalog combines data from multiple HST programs with different exposure times, we additionally tested whether field-dependent depth variations could influence the comparison between observed and simulated densities. In particular, the pointing containing NGC~4889 (field 3-3) is deeper than the surrounding ACS fields. Assuming background-limited noise, for which the limiting flux scales as $\propto t^{-1/2}$, the exposure-time ratio implies a depth offset of $\Delta m \simeq 1.07$ mag relative to the adjacent fields\footnote{In the background-limited regime, appropriate for deep ACS/WFC imaging (e.g. \citealt{Sirianni2005}), the noise is dominated by sky fluctuations rather than read noise, such that $\psub{f}{lim} \propto t^{-1/2}$ and $\Delta m = 1.25 \log_{10}\!\left(\frac{t_1}{t_2}\right)$.}. An empirical estimate based on background RMS was explored but found to be unreliable due to contamination from extended galaxy light in the overlap regions.

To account for this pointing difference we modeled the completeness in two stages representing both detection and GC-candidate selection. The overall inclusion probability was written as $\psub{p}{inc}(m)=\psub{p}{det}(m)\,\psub{p}{sel}(m)$ and folded through the adopted GCLF to derive a pointing-dependent effective recovery fraction. These recovery fractions were incorporated into the effective survey area used when constructing the Voronoi density grids. This procedure ensures that both the simulated and observed density maps are compared under a consistent completeness model across the full mosaic.

After completion of the simulation runs, the observed grid density was compared to the simulation density to give the residual
\begin{align}
    \Delta\Sigma &= \rho_\text{obs}-\langle\rho\rangle
    \intertext{and the significance, $S$, was determined using this, the observed data Poisson error and the simulation standard deviation, via}
    S(x,y) &= \frac{\Delta\Sigma }{\sqrt{\sigma^2_\text{obs}+\sigma^2_\rho}} \label{eq:2d_signif}
\end{align}
For clarity the significance was not overlaid on background HST F814W images. 

Physically, the significance is the residual of the surface density of GCs relative to the combined uncertainty (\ie observed Poisson uncertainty and the model variance). In practical terms, $S>0$ indicates a local excess relative to the model prediction, while $S<0$ indicates a deficit. Such features may arise from environmental processes including accretion, stripping, or the presence of intracluster GC populations. We also note that, under an approximate Gaussian-noise interpretation, $\abs{S}\geq2$ corresponds to a deviation at roughly the $2\sigma$ level.

We additionally tested whether the residual structures could arise from instrumental effects associated with the ACS pointing pattern or field-dependent depth variations. To do this we compared the total observed and simulated GC densities on a per-pointing basis across the mosaic. The resulting observed-to-simulated density ratios show that the strongest excesses occur in the fields centered on the dominant cluster galaxies (\eg NGC~4874, NGC~4889, and IC~4051), while several other pointings show ratios close to unity. Because the largest excess does not coincide with the deepest pointing, these tests indicate that the residual structures are not driven by exposure-depth variations or ACS field boundaries.

The resulting significance map highlights spatial deviations between the observed GC distribution and the model predictions derived from the galaxy population.

In \cref{fig:Coma_cluster_2D_Vor_density}, we show the residual significance of the projected 2D density map for the observed portion of the Coma cluster. From this full cluster overview, we note there are many galaxies which show either a deficit compared to that predicted from \SN, and additionally galaxies which exhibit an azimuthal asymmetry in respect of their GC distribution. We thus proceed to investigate this further for the galaxies in our sample.

We present a fully detailed example of this process in \cref{fig:Coma_NGC 4874_2D_Vor_density}. In this figure, centered on the BCG NGC~4874, we show the plots derived from the 2D Voronoi density map for the observed GC data (\cref{fig:Coma_NGC 4874_2D_Vor_obs}), simulated GC field (\cref{fig:Coma_NGC 4874_2D_Vor_sim}), the residual between the two (\cref{fig:Coma_NGC 4874_2D_Vor_res}) and the significance (\cref{fig:Coma_NGC 4874_2D_Vor_sig}). The scales of the observed and simulated data are the same to allow direct comparison. The blue ring around the central galaxy denotes the effective radius used for the GCS (\ReffGCS) used for the generation of the model data. As previously, major galaxies (\ie NGC/IC catalog) were labeled and marked with a red cross. Other, unlabeled galaxies ($\MV\qty{<-19.6}{\mag}$) for which the GC populations were also simulated, are marked with a green cross. For reference, we overlay the HST/ACS pointings in green \citep[GO 10861, GO 11711, GO 12918;][respectively]{Carter_2008, Cho_2016, Harris_2017b}.

We make several general observations on the basis of the panels in \cref{fig:Coma_NGC 4874_2D_Vor_density}. Firstly, as noted by authors in the past \citep[\eg][]{Marin-Franch_2002, Harris_2009, Peng_2011, Cho_2016, Madrid_2018}, the agglomeration of GCs around NGC~4874 is more dense than would be predicted merely from the magnitude of the galaxy. However, there are other subtle nuances which become clear after this analysis. For example, there are many voids in the GC distribution, such as that in the south-east corner of the image, around NGC~4876, and NGC~4872 to the south of NGC~4874, both of which were flagged in Section \ref{sec:1D_voronoi} as galaxies with a potential deficit of GCs. There are also prominent asymmetries such as the excess around NGC~4873 to the north-west of the BCG, where the excess is concentrated toward the gravity well of the BCG. This asymmetry and that of the unlabeled galaxy SDSS J125935.18+275605.0, (green cross due south of NGC~4874) were identified by \citet{Cho_2016}, which validates this method as a diagnostic.

\subsection{Galaxies with deficit GCS population}
For the other galaxies noted in Section \ref{sec:Coma_gal_GCSF} as potentially having deficit GC populations, we present \cref{fig:Coma_other_gals_2D_Vor_density}, where we show just the final significance plots to illustrate the GC deficit we have identified surrounding these galaxies. These include NGC~4876 \cref{fig:Coma_NGC 4876_2D_Vor_sig}, NGC~4883 \cref{fig:Coma_NGC 4883_2D_Vor_sig}, NGC~4908 \cref{fig:Coma_NGC 4908_2D_Vor_sig}, and IC~3973 \cref{fig:Coma_IC 3973_2D_Vor_sig}. The parameters for these plots are the same as those discussed above.  

The deficit of the GCSs compared to that predicted using the GC specific frequency relation is especially distinct around the four galaxies shown in \cref{fig:Coma_other_gals_2D_Vor_density}. However, visually only NGC~4908 (\cref{fig:Coma_NGC 4908_2D_Vor_sig}) sits in an isolated trough in the GC distribution, with only minor asymmetries evident. The other three galaxies included in \cref{fig:Coma_other_gals_2D_Vor_density} (NGC~4876 \cref{fig:Coma_NGC 4876_2D_Vor_sig}, NGC~4883 \cref{fig:Coma_NGC 4883_2D_Vor_sig}, and IC~3973 \cref{fig:Coma_IC 3973_2D_Vor_sig}) are located in depressions in the GC distribution which exhibit far more complex structure and large scale asymmetries.

\section{Azimuthal Asymmetries and Evidence for Environmental Stripping}\label{sec:az_symmetry}
Azimuthal asymmetry in a GCS can provide evidence of past environmental interactions such as tidal stripping, mergers, or accretion. We therefore test the GC candidate distributions around the galaxies in our sample for departures from circular symmetry. We include our suspect deficit GC population galaxies in this for comparison with the remaining galaxies ($\MV<-19.6$ mag) in our list. We do, however, exclude the BCGs, any galaxies in their immediate vicinity ($r\!\lesssim\!\ReffGCS$) and those within a projected distance of $\lesssim\!\ang{;;60}$ of the edge of the HST pointing region. 

\subsection{Circular uniformity Rayleigh test}
To test specifically for dipole-like departures from circular symmetry, we first apply a Rayleigh test to the GC population within a projected radius of $\ang;;60$ around each galaxy in our sample. The Rayleigh test is sensitive to a preferred angular direction in the GC distribution, as expected if environmental stripping or displacement produces a one-sided excess or deficit of clusters. We use the \texttt{rayleightest} implementation in the \texttt{astropy.stats} package, which returns a Rayleigh $p$-value under the null hypothesis of circular uniformity, \ie
\begin{multline}
    \psub{p}{Rayleigh} = \exp(-Z)\left[1+\frac{(2Z-Z^2)}{4n}-\right.\\
    \left.\frac{24Z-132Z^2+76Z^3-9Z^4}{288n^2}\right]
\end{multline}
where the test statistic $Z$ is
\begin{align}
    Z = n\Bar{R}^2 &= \frac{R^2}{n}
    \intertext{from the resultant vector length, $R$, and mean resultant vector, $\Bar{R}$, given by}
    R = n\sqrt{\Bar{C}^2 + \Bar{S}^2}  \quad&\text{and}\quad \Bar{R}=\frac{R}{n}
    \intertext{from the mean $\cos\theta$ and $\sin\theta$ for each galaxy GC population}
    \Bar{C} = \frac{1}{n}\sum_{i=1}^n\cos{\theta_i} \quad&\text{and}\quad \Bar{S} = \frac{1}{n}\sum_{i=1}^n\sin{\theta_i}
    \intertext{and from this we obtain the mean resultant vector $\overline{PA}$ using}
    \Bar{\theta} &= \tan^{-1}\frac{\Bar{S}}{\Bar{C}}
\end{align}

Observationally, each GC contributes a unit vector defined by its projected position angle (PA), $\theta_i$, relative to the galaxy center. The resultant vector length $R$ is the magnitude of the sum of these vectors, and its normalized form $\Bar{R}$ measures the degree to which the GC system exhibits a preferred orientation on the sky rather than an isotropic angular distribution. The Rayleigh statistic is thus sensitive to a global preferred angular direction in the GC population by using the full set of position angles.

\subsection{Hemisphere binomial / K-S test}
For our second test, we perform complementary two-hemisphere binomial and Kolmogorov–Smirnov (K-S) tests, comparing the counts and radial distributions for asymmetry in opposite hemispheres of the same galaxy. For these tests we use the \texttt{scipy.stats} package implementation of \texttt{binomtest} and \texttt{ks\_2samp}, respectively. 

\renewcommand{\arraystretch}{1.2}
\begin{deluxetable*}{lccccccccl}
\tabletypesize{\scriptsize}
\tablecaption{GCS asymmetry test results for selected Coma galaxies.  Ordered by increasing \psub{p}{Rayleigh}.
\label{tab:asymmetry_results}}
\tablehead{
	\colhead{Galaxy} & \colhead{\psub{N}{GC}} & \multicolumn{3}{c}{Rayleigh} & \multicolumn{2}{c}{Hemisphere} & \multicolumn{2}{c}{Quadrant} & \colhead{Comments} \\ 
	 &  & \colhead{$\log_{10} p$} & \colhead{$\overline{PA}$} & \colhead{$\overline{R}$} & \colhead{$\log_{10} \psub{p}{binom}$} & \colhead{$\log_{10} \psub{p}{K-S}$} & \colhead{$\chi^2$} & \colhead{$\log_{10} p$} &  
}
\startdata
	IC 3998 & 433 & $\num{-18.61}^{+\num{3.60}}_{-\num{4.01}}$ & $266$ & $0.31$ & $\num{-14.83}^{+\num{3.29}}_{-\num{3.75}}$ & $\num{-5.30}^{+\num{1.91}}_{-\num{2.18}}$ & $65.48$ & $\num{-13.91}$ & \asymcomment{Extremely strong asymmetry - clear directional excess \ang{\sim266;;}, with high $\overline{R}$ (0.31). Likely real structure (\eg tidal or cluster-facing truncation).} \\
	NGC 4876 & 116 & $\num{-12.92}^{+\num{2.67}}_{-\num{3.14}}$ & $326$ & $0.50$ & $\num{-10.44}^{+\num{2.76}}_{-\num{2.62}}$ & $\num{-2.14}^{+\num{0.97}}_{-\num{1.28}}$ & $76.21$ & $\num{-15.95}$ & \asymcomment{Strongly asymmetric; dipole at $\!\sim\!\ang{326;;}$. Extremely significant.} \\
	NGC 4875 & 230 & $\num{-5.04}^{+\num{1.80}}_{-\num{2.13}}$ & $333$ & $0.22$ & $\num{-6.53}^{+\num{2.29}}_{-\num{2.40}}$ & $\num{-4.37}^{+\num{1.64}}_{-\num{2.45}}$ & $19.70$ & $\num{-4.08}^{+\num{1.46}}_{-\num{1.95}}$ & \asymcomment{Clear lopsidedness; likely GC deficit or tidal direction toward cluster potential.} \\
	NGC 4883 & 104 & $\num{-3.66}^{+\num{1.52}}_{-\num{1.91}}$ & $220$ & $0.28$ & $\num{-2.66}^{+\num{1.03}}_{-\num{1.68}}$ & $\num{-0.58}^{+\num{0.33}}_{-\num{0.68}}$ & $20.92$ & $\num{-4.11}^{+\num{1.71}}_{-\num{2.45}}$ & \asymcomment{Angular asymmetry but no significant radial difference (K-S high); consistent with uneven azimuthal coverage or partial truncation.} \\
	IC 4042 & 155 & $\num{-2.71}^{+\num{1.17}}_{-\num{1.72}}$ & $147$ & $0.19$ & $\num{-2.66}^{+\num{1.23}}_{-\num{1.94}}$ & $\num{-1.74}^{+\num{0.87}}_{-\num{1.31}}$ & $8.28$ & $\num{-1.82}^{+\num{0.95}}_{-\num{1.35}}$ & \asymcomment{Mild-to-moderate asymmetry.} \\
	NGC 4908 & 218 & $\num{-2.29}^{+\num{1.22}}_{-\num{1.61}}$ & $209$ & $0.15$ & $\num{-2.27}^{+\num{1.23}}_{-\num{1.71}}$ & $\num{-6.70}^{+\num{2.34}}_{-\num{2.58}}$ & $9.12$ & $\num{-2.01}^{+\num{1.00}}_{-\num{1.37}}$ & \asymcomment{K-S p-value very low — radial deficit on one side ($\Delta r\!\sim\!\ang{;;19}$); strong candidate for environmental stripping.} \\
	NGC 4869 & 403 & $\num{-1.98}^{+\num{1.02}}_{-\num{1.40}}$ & $41$ & $0.10$ & $\num{-2.15}^{+\num{1.03}}_{-\num{1.51}}$ & $\num{-0.60}^{+\num{0.37}}_{-\num{0.78}}$ & $5.11$ & $\num{-1.14}^{+\num{0.72}}_{-\num{1.13}}$ & \asymcomment{Weak-to-moderate signal — maybe mild elongation, not clearly significant.} \\
	IC 4041 & 117 & $\num{-1.56}^{+\num{0.88}}_{-\num{1.26}}$ & $346$ & $0.17$ & $\num{-1.80}^{+\num{1.09}}_{-\num{1.60}}$ & $\num{-1.80}^{+\num{1.19}}_{-\num{2.08}}$ & $2.15$ & $\num{-0.62}^{+\num{0.42}}_{-\num{0.78}}$ & \asymcomment{Mild asymmetry (mostly hemisphere counts); intrinsic ellipticity or projection of an elongated GCS, unlikely environmental.} \\
	IC 4045 & 170 & $\num{-1.51}^{+\num{0.91}}_{-\num{1.45}}$ & $94$ & $0.13$ & $\num{-1.59}^{+\num{0.87}}_{-\num{1.21}}$ & $\num{-0.88}^{+\num{0.47}}_{-\num{0.75}}$ & $10.61$ & $\num{-2.22}^{+\num{1.14}}_{-\num{1.52}}$ & \asymcomment{Weak anisotropy (slight elongation); intrinsic flattening more likely than stripping.} \\
	IC 4040 & 98 & $\num{-1.24}^{+\num{0.74}}_{-\num{1.05}}$ & $260$ & $0.15$ & $\num{-1.26}^{+\num{0.69}}_{-\num{1.23}}$ & $\num{-0.85}^{+\num{0.49}}_{-\num{0.99}}$ & $4.12$ & $\num{-0.97}^{+\num{0.62}}_{-\num{1.08}}$ & \asymcomment{No significant asymmetry; undisturbed halo.} \\
	IC 4026 & 108 & $\num{-1.18}^{+\num{0.74}}_{-\num{1.32}}$ & $355$ & $0.14$ & $\num{-0.99}^{+\num{0.69}}_{-\num{1.05}}$ & $\num{-0.96}^{+\num{0.56}}_{-\num{1.07}}$ & $6.00$ & $\num{-1.42}^{+\num{0.93}}_{-\num{1.21}}$ & \asymcomment{Mild hints of asymmetry; likely isotropic.} \\
	IC 3973 & 89 & $\num{-0.80}^{+\num{0.58}}_{-\num{1.01}}$ & $323$ & $0.13$ & $\num{-1.05}^{+\num{0.77}}_{-\num{1.51}}$ & $\num{-0.85}^{+\num{0.51}}_{-\num{0.99}}$ & $1.56$ & $\num{-0.52}^{+\num{0.37}}_{-\num{0.76}}$ & \asymcomment{Weak angular imbalance, no radial difference.} \\
	IC 4030 & 45 & $\num{-0.76}^{+\num{0.56}}_{-\num{1.01}}$ & $333$ & $0.16$ & $\num{-0.87}^{+\num{0.61}}_{-\num{0.92}}$ & $\num{-0.69}^{+\num{0.43}}_{-\num{0.81}}$ & $3.80$ & $\num{-0.97}^{+\num{0.61}}_{-\num{0.87}}$ & \asymcomment{No real asymmetry.} \\
	IC 3976 & 74 & $\num{-0.77}^{+\num{0.53}}_{-\num{0.86}}$ & $32$ & $0.12$ & $\num{-0.89}^{+\num{0.63}}_{-\num{1.26}}$ & $\num{-1.54}^{+\num{0.91}}_{-\num{1.49}}$ & $5.46$ & $\num{-1.31}^{+\num{0.71}}_{-\num{0.95}}$ & \asymcomment{Angularly uniform but one-sided radial difference; marginal local environment effect.} \\
	IC 4033 & 58 & $\num{-0.66}^{+\num{0.47}}_{-\num{0.85}}$ & $194$ & $0.13$ & $\num{-0.83}^{+\num{0.67}}_{-\num{0.78}}$ & $\num{-1.00}^{+\num{0.64}}_{-\num{1.31}}$ & $3.38$ & $\num{-0.87}^{+\num{0.63}}_{-\num{0.97}}$ & \asymcomment{Isotropic.} \\
	IC 4021 & 80 & $\num{-0.60}^{+\num{0.43}}_{-\num{0.79}}$ & $166$ & $0.10$ & $\num{-0.84}^{+\num{0.60}}_{-\num{0.90}}$ & $\num{-0.88}^{+\num{0.55}}_{-\num{0.89}}$ & $5.50$ & $\num{-1.24}^{+\num{0.71}}_{-\num{1.07}}$ & \asymcomment{Isotropic.} \\
	NGC 4906 & 131 & $\num{-0.31}^{+\num{0.24}}_{-\num{0.55}}$ & $28$ & $0.01$ & $\num{-0.53}^{+\num{0.39}}_{-\num{0.74}}$ & $\num{-1.37}^{+\num{0.91}}_{-\num{2.07}}$ & $4.85$ & $\num{-1.14}^{+\num{0.70}}_{-\num{1.06}}$ & \asymcomment{Completely symmetric distribution; good example unperturbed / control galaxy.} \\
\enddata
\tablecomments{All $p$-values are reported as $\log_{10} p$ with 68\% bootstrap intervals. Columns (1) - Galaxy ID; (2) GC count within \ang{;;60}; (3) Rayleigh $p$ - Detection of preferred direction (low $p$ values indicate strong directional clustering); (4) $\overline{PA}$ - Mean resultant direction (east of north) ; (5) $\overline{R}$ - Mean, normalized resultant vector magnitude; (6) Hemisphere binomial $p$ - Count imbalance between opposite hemispheres; (7) Hemisphere K-S $p$ - Radial distance differences; (8) Quadrant $\chi^2$ - Test for general isotropy; (9) Quadrant $p$ - Probability that such a distribution arises; (10) Comments.}
\end{deluxetable*}

Observationally, these hemisphere tests probe a different class of asymmetry from the Rayleigh test. The hemisphere binomial test measures net count imbalances between opposing halves of a galaxy, independent of the detailed angular distribution within each hemisphere. The K–S test extends this comparison by testing the radial distributions of GCs between hemispheres, allowing the identification of asymmetric radial structure even if the total GC counts are comparable. Together, these tests distinguish between coherent angular alignment, gross lopsidedness, and one-sided radial distortions in the GC systems.

\subsection{Quadrant \texorpdfstring{$\chi^2$}{chi2} test}
Finally, we implement a $\chi^2$ test on quadrants to test general anisotropy, \ie elongation or clumping, of the GCS for each of the galaxies in our sample. With the distribution divided into quadrant bins ($k=4$), we have
\begin{align}
    \chi^2 &= \sum_{i=1}^k\frac{(N_i-\Bar{N})^2}{\Bar{N}}
    \intertext{where $N_i$ is the number of GCs in bin $i$, and}
    \Bar{N} &= \frac{1}{k}\sum_{i=1}^k N_i
    \intertext{and with the degrees of freedom, $\nu = k-1$, the $p$-value is then}
    \psub{p}{quad} &= 1 - F_{\chi^2}\left(\chi^2;\nu\right)
\end{align}
where $F_{\chi^2}$ is the $\chi^2$ cumulative distribution function.

\subsection{Analysis}
We fix a common search radius ($\psub{r}{max}\!=\!\ang{;;60}$) for all the galaxies and select the GC within that radius. We compute the \texttt{pa\_deg} and \texttt{sep\_arcsec} and apply the Rayleigh dipole test using this data. We obtain a preferred angle and mean resultant length, $\overline{R}$, the former of which is used to center the split for the opposite hemispheres on the K-S test. We then perform a binomial test on the counts and a two-sample K-S test on the radial distances. We finally perform a quadrant $\chi^2$ test to check for broader non-uniformity. The confidence interval on the $p$-values is determined through bootstrap resampling (1000 iterations). The result for the 17 galaxies in our subsample are presented in Table \ref{tab:asymmetry_results}. 

\begin{figure}[ht]
    \centering
    \includegraphics[width=1\linewidth]{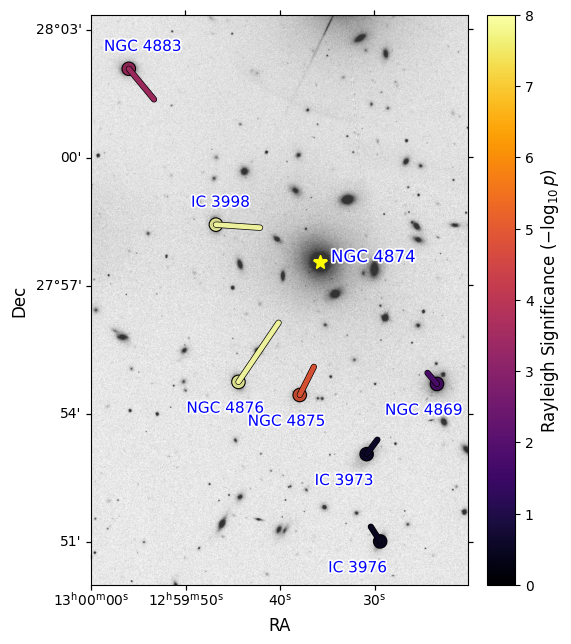}
    \caption{GCS asymmetry vectors for galaxies from our sample, around the BCG NGC~4874. The length of the vector scales with the mean resultant vector, $\Bar{R}$, and the color represents the Rayleigh $p$-value probability under isotropy, as described in the text. The alignment of the directional GC excesses with the BCG NGC~4874 is clear.}
    \label{fig:NGC_4874_GCS_Asymm}
\end{figure}

In general, we note that a low $p$ coupled with a high $\overline{R}$ is suggestive of directional asymmetry, indicative of possible stripping/truncation. Additionally, a low quadrant $p$ with a moderate $\overline{R}$ more indicative of GCS elongation, more likely an intrinsic flattening. Finally, a high $p$ signature everywhere is indicative of an isotropic GC population, suggesting any deficit against predictions is intrinsic. 

A potential concern is that spatial variations in detection efficiency relative to the host galaxy light could introduce artificial asymmetries. In particular, reduced completeness along the major axis of highly elliptical galaxies could produce distorted structures when compared with circular models. To test this possibility we compared the measured asymmetry axes with the stellar major-axis orientations of the host galaxies. The offsets between the two axes are broadly distributed with a median offset of $\ang{\approx 40;;}$. A K-S test shows that the distribution of offsets is statistically consistent with a uniform distribution ($p=0.99$ for the full sample), indicating no evidence for systematic alignment between the asymmetry axes and galaxy major axes. Restricting the analysis to galaxies with statistically significant asymmetries yields similar results ($p=0.69$). These tests suggest that the observed asymmetries are unlikely to arise from completeness variations associated with galaxy isophotes.

From this analysis, we find evidence for strong asymmetries in IC~3998, NGC~4876 and NGC~4875. These are real, directional GC excesses. The directions, $\overline{PA}=[\ang{266;;},\ang{326;;},\ang{333;;}]$, respectively, align directly with the BCG NGC~4874, as illustrated clearly in the cutout \cref{fig:NGC_4874_GCS_Asymm}. Their location and the directional excesses are suggestive of tidal stripping, ram-pressure removal, or GC migration into the intracluster medium. 

We also find moderate asymmetries associated with NGC~4883, IC~4042, NGC~4908. These galaxies have asymmetric GCS envelopes or have experienced partial depletion on one side. There is angular asymmetry associated with NGC~4883, but we find no significant radial difference (K-S high) and this is consistent with uneven azimuthal coverage or partial truncation. The K-S signal associated with NGC~4908 ($\log_{10}p\!\sim\!\num{-6.7}$, $\Delta r\!\sim\!\ang{;;19}$) implies a different radial depth between hemispheres, which is a signature of truncation or in-fall. 

The remaining galaxies show only mild or no significant asymmetries, suggesting minimal past interaction, and thus any deficit in the GCS is most likely intrinsic, but could also be the result of a symmetric truncation process, such as galaxy and its GCS performing multiple orbits around the center of the BCG. 

Taken together, the Rayleigh, hemisphere, and quadrant tests provide consistent evidence that several GC systems in the central region of Coma exhibit statistically significant azimuthal asymmetries. These asymmetries are not correlated with the stellar major-axis orientations of the host galaxies and therefore are unlikely to arise from observational completeness effects. Instead, their preferential alignment toward the cluster center supports an interpretation in which environmental processes - such as tidal stripping, GC migration into the intracluster medium, or asymmetric truncation of GC halos - play an important role in shaping the GC populations of galaxies in the dense core of the Coma cluster.

\section{Discussion}\label{sec:p2_results_discuss}

In the above analysis we have demonstrated that several galaxies in the observed region of the Coma cluster host fewer GCs than expected from their luminosities. Broadly, these deficits may arise either from intrinsic differences in GC formation efficiency or from environmental and evolutionary processes acting within the cluster environment. We briefly consider both possibilities below.

\subsection{Intrinsic Deficit}
An intrinsic deficit, refers to a galaxy’s whose GC population is low, relative to its stellar luminosity or mass, but native to its formation history rather than the result of subsequent environmental depletion. Such a deficit could arise from unusually low GC formation efficiency at early times, a mismatch between halo mass and stellar luminosity, or early truncation of GC formation.

Low cluster formation efficiency (CFE) may result from poor gas conditions early in the formation of a galaxy with its GC population \citep[\eg][]{Kruijssen_2014, Li_2018, Kruijssen_2026}. For example, low-mass galaxies (with shallow potentials) tend to have lower gas surface densities and pressures. Authors such as \citet{Kruijssen_2012} and \citet{Adamo_2015} show this leads to a lower CFE, although we note this is unlikely the case for the high-mass galaxies we consider in our Coma dataset. Additionally, strong feedback and reionization suppress cooling pathways \citep[\eg][]{Borrow_2023}, especially in shallow potentials, and modeling has shown a resulting reduction in the formation of clusters \citep[\eg][]{Howard_2017}. The CFE \citep{Kruijssen_2012} along with environmental observational work \citep[\eg][]{Adamo_2015, Adamo_2020} has shown that lower density, pressure, or more disruptive environments yield a smaller fraction of stars in bound clusters.

A mismatch between the galaxy halo mass and its luminosity is a possibility if a recent star formation episode boosted the luminosity without increasing the halo mass \citep[\eg][]{Bell_2001, Sparre_2017}. In this event, the GC population, predicted on the stellar luminosity will be over estimated. The sensitivity of \SN to luminosity changes has been discussed in the past \citep[\eg][]{Peng_2008, Liu_2019}, while others note that the GC population more closely traces halo mass \citep[\eg][]{Hudson_2014, Harris_2015} and thus luminosity-based predictions can overestimate GC counts in recently brightened systems.

During GC formation, early truncation may occur if gas flow is quenched \citep[\eg][]{Kruijssen_2014, Forbes_2018a}, or if pre-processing occurs in groups before cluster in-fall can occur, cutting off a metal-rich (red) GC phase \citep[\eg][]{Pfeffer_2023, Chen_2025}. This would yield a lower total GC population for the host galaxy, but also a metal-poor (blue) dominated system as theorized by \citet{Kruijssen_2014, Forbes_2018a} and observed by \citet{Janssens_2022}.

\subsection{Environment/Evolution}
In a cluster environment, we would expect the removal or redistribution of GC populations as a result of interactions between cluster members. However, these interactions may take many forms, resulting in different end effects. These interactions include tidal stripping by the BCG potential, multiple close-encounters, dynamical friction, enhanced disruption and dark matter stripping. 

Tidal stripping of a host's GCS may occur by interaction with the BCG potential \citep[\eg][]{Bekki_2003, Peng_2011, Ramos_2015}. Tidal stripping occurs on the host outer halo first, due to the spatially extended nature of GCSs, and thus GC counts are lowered within a fixed aperture truncating the radial profile \citep[\eg][]{Bekki_2003, Hudson_2014}. Stripped GCs are redistributed to the intracluster environment \citep[\eg][]{Durrell_2014, Ahvazi_2024}, and as demonstrated for Coma by \citet{Peng_2011}. In our case we note that galaxies closer to the BCG NGC~4874 (\ie IC~3998 and NGC~4876) show the strongest asymmetries, suggesting a cluster centric dependence in stripping efficiency. Similar extended blue globular cluster components, interpreted as intracluster or stripped populations, have recently been identified in the Hydra I cluster \citep{Lohmann_2026}, supporting the broader environmental picture.

Cumulative tidal shocks from repeated high-speed encounters can heat and strip clusters, particularly in the outer halo, which manifests as asymmetric deficits towards the cluster center \citep[\eg][]{Smith_2013, Blana_2025}.

In intermediate mass galaxy hosts, massive GCs can sink and merge into a nuclear star cluster in the UCD size/mass space \citep[\eg][]{Neumayer_2020, Fahrion_2022}. In larger hosts, such as those in our Coma dataset, the tidally stripped remnants of accreted dwarfs may also appear similar to these UCD-like objects \citep[\eg][]{Norris_2011, Khoperskov_2023}. Both of these channels will siphon GCs from the host bound GC population.

Enhanced disruption may be caused by a combination of strong external tidal field in the cluster potential \citep[\eg][]{Baumgardt_2003, Madrid_2017} and internal shocks, \eg\ passages through a dense disk, or close to a bulge \citep[\eg][]{Gnedin_1998, Webb_2019}. These processes accelerate two-body evaporation \citep[\eg][]{Baumgardt_2003, Madrid_2017}, and in this case the deficit will be stronger at small radii in low-mass clusters \citep[\eg][]{Gnedin_1998, Webb_2013}; see also \citet{Forbes_2018a} for an environment-dependent disruption review.

If the dark matter halo of a host has been heavily stripped, then the present day tidal radius will be small \citep[\eg][]{Ogiya_2022}. In this situation, the surviving GCS will be compact and exhibit a deficit relative to that predicted by the unstripped stellar light \citep[\eg][]{Hudson_2014}.

Thus, the combination of GC deficits and azimuthal asymmetry signatures provides a tracer of the evolutionary state of galaxies within the Coma cluster environment. We suggest that IC~3973 and IC~4040 are more likely intrinsically low-GC galaxies, potentially reflecting early quenching or pre-processing prior to cluster infall. In contrast, IC~3998, NGC~4876, NGC~4883, and NGC~4908 show clear evidence of asymmetric or truncated GC systems, consistent with ongoing or recent environmental stripping. Any GCs removed through these processes likely contribute to the intracluster GC population surrounding the BCGs.

This evidence also allows us to argue that the observed scatter in the GC specific frequency within clusters may be dominated by environmental erosion rather than any intrinsic formation efficiency scatter. This aligns with the GCS frameworks proposed by \citet{Peng_2008} and \citet{Hudson_2014}.

\section{Conclusions}\label{sec:p2_conclusion}

We carried out our analysis for the brighter galaxies within the central \qty{1}{\mega\parsec} of the Coma cluster, incorporating hundreds of galaxies surrounding the dual BCGs, NGC~4874 and NGC~4889. From an initial comparison between the background-corrected observed GCS counts and those predicted from the specific frequency \SN\ based on visual magnitude \MV, we identified a number of statistically significant outliers exhibiting anomalously low GCS counts. Further analysis of the radial GC surface-density profiles around the BCGs, using a detailed Voronoi tessellation, revealed evidence for local imprints of galaxy GCS populations on the surrounding density field, as well as radial offsets in some cases. Other galaxies, however, showed no clear imprint on the radial profile.

We extended this Voronoi density analysis of our candidate GC dataset to a 2D map projected over the observed region. By simulating the GCS associated with our sample galaxies on the basis of isolated specific frequency prediction, we obtained a residual difference and determine the significance of this from the observed data Poisson uncertainty and the simulation standard deviation. The significance map showed regions where there is a significant excess, \eg around NGC~4874 and IC~4051, as seen previously. It also revealed regions of deficit relative to the predictions around several galaxies, notably NGC~4876, NGC~4883, NGC~4908, and IC~3973.

We verified that these structures are not driven by variations in observational depth or by instrumental field boundaries, and that the detected asymmetries are not aligned with the stellar major axes of the host galaxies, indicating that they are unlikely to arise from completeness effects.

To explore the nature of these deficits, we performed a series of azimuthal symmetry tests on the GCS around 17 of the sample galaxies. Three galaxies (IC~3998, NGC~4876 and NGC~4875) show strong asymmetries ($\log\psub{p}{Rayleigh}<-5$) toward the BCG NGC~4874, which is strongly suggesting past interaction. This also indicates the deficit noted around NGC~4876 is likely related to such an interaction. We find moderate asymmetries ($-5<\log\psub{p}{Rayleigh}<-2$) around three further galaxies (NGC~4883, IC~4042, NGC~4908). A lack of radial difference around NGC~4883 is consistent with partial truncation, where again, the PA (\ang{220;;}) is directed toward the BCG NGC~4874. Conversely, the one-sided radial deficit around NGC~4908 makes it a strong candidate for environmental stripping. A fourth galaxy, NGC~4869, shows only mild asymmetry, but we note that the excess PA (\ang{41;;}) is again in the direction of the BCG NGC~4874.

The remaining galaxies show little or no asymmetries, which is consistent with the possibility that their deficits are intrinsic, rather than environmentally induced. These include IC~3973, IC~3976, IC~4040, and IC~4045.

Together, these results suggest that Coma’s dense environment has measurably altered the GCSs of some of its member galaxies, producing evidence for tidal truncation and stripping, while other systems may instead host intrinsically under-populated GC systems. Future work comparing GC populations across clusters such as Fornax may further illuminate environmental differences. Future wide-field imaging and spectroscopic studies will be important for determining whether these deficit systems contribute directly to the intracluster GC population, and for further constraining the role of environmental stripping in shaping GC systems in dense cluster environments.


\begin{acknowledgments}

We thank the anonymous referee for a prompt and constructive report that helped us to improve this paper.

We would like to thank Carlos J. Donzelli (Instituto de Astronomía Teórica y Experimental (IATE), CONICET – U. Nacional de Córdoba, X5000BGR Córdoba, Argentina) for sharing his insights and experience on the galaxy light modeling in this paper.

We thank Blanca O. Garcia and Katrina Martinez, at The University of Texas Rio Grande Valley, for their help and support for this project.

Based on observations made with the NASA/ESA Hubble Space Telescope, obtained at the Space Telescope Science Institute, which is operated by the Association of Universities for Research in Astronomy, Inc., under NASA contract NAS5-26555. These observations are associated with programs GO 10861, 11711, 12918. 

This research has made use of the NASA Astrophysics Data System Bibliographic services (ADS), funded by NASA under Cooperative Agreement 80NSS-C21M00561. 

This research has made use of the SIMBAD database, operated at Centre de Donn\'{e}es astronomiques de Strasbourg (CDS), Strasbourg, France \citep{Wenger_2000}.

This research has made use of the NASA/IPAC Extragalactic Database (NED), which is funded by the National Aeronautics and Space Administration and operated by the California Institute of Technology.

Funding for the Sloan Digital Sky Survey IV has been provided by the Alfred P. Sloan Foundation, the U.S. Department of Energy Office of Science, and the Participating Institutions.  SDSS-IV acknowledges support and resources from the Center for High Performance Computing  at the University of Utah. The SDSS website is \url{https://www.sdss4.org/}.

This work is supported by the National Science Foundation under Cooperative Agreement PHY-2019786 (The NSF AI Institute for Artificial Intelligence and Fundamental Interactions, \url{http://iaifi.org/}).
\end{acknowledgments}

All the {\it HST} data used in this paper can be found in MAST: \dataset[10.17909/T97P46]{http://dx.doi.org/10.17909/T97P46} \citep{STSCI_MAST}.

\software{Astropy \citep{astropy:2013, astropy:2018, astropy:2022}, astroquery \citep{Ginsburg_2019}, matplotlib \citep{Hunter_2007}, numpy \citep{harris_c_r_2020}, phoutils \citep{Bradley_2025}, pandas \citep{mckinney_2010}, scipy \citep{scipy_2020}, statsmodels, \citep{statsmodels_2010}}.

\appendix

\section{GC specific frequency - Non-detections}\label{app:GCSF-robustness}

\begin{deluxetable}{lccccccccc}
\tablecaption{Galaxy GC specific frequencies: Upper limits\label{tab:raw_GCSF_data_nondet_app}}
\tablehead{
    \colhead{Galaxy} & \colhead{MV (mag)} & \colhead{$\psub{N}{corr}$} & \colhead{non-det UL $1\sigma$} & \colhead{edge\_flag} & \colhead{nearest BCG} & \colhead{BCG dist. (")}
}
\startdata
NGC 4898A & -20.92 & 16(12) & 0.12 & edge & NGC 4889 & 149.2 \\
NGC 4886 & -20.78 & 986(41) & 5.01 &  & NGC 4889 & 61.5 \\
Z 160-A14 & -20.68 & -81(7) & 0.03 &  & NGC 4874 & 510.5 \\
NGC 4883 & -20.66 & 11(9) & 0.11 &  & NGC 4889 & 262.2 \\
LEDA 44708 & -20.55 & 4736(80) & 29.06 &  & NGC 4889 & 26.8 \\
NGC 4876 & -20.51 & -13(7) & 0.04 &  & NGC 4874 & 204.1 \\
NGC 4872 & -20.42 & -10(18) & 0.12 &  & NGC 4874 & 49.3 \\
IC 4026 & -20.37 & 12(9) & 0.16 &  & NGC 4889 & 313.2 \\
IC 4041 & -20.28 & -16(10) & 0.08 &  & NGC 4889 & 439.6 \\
Z 160-261 & -20.26 & 0(0) & 0.00 &  & NGC 4889 & 732.7 \\
IC 4012 & -20.10 & 10(4) & 0.12 & edge & NGC 4889 & 365.5 \\
NGC 4894 & -20.08 & -93(14) & 0.13 & edge & NGC 4889 & 116.7 \\
IC 4021 & -20.07 & 16(5) & 0.20 &  & NGC 4889 & 247.6 \\
NGC 4898B & -20.05 & 19(11) & 0.29 & edge & NGC 4889 & 151.5 \\
IC 3998 & -20.03 & -27(14) & 0.14 &  & NGC 4874 & 155.8 \\
IC 4040 & -19.96 & 12(8) & 0.21 &  & NGC 4889 & 490.3 \\
IC 4033 & -19.95 & -5(7) & 0.07 &  & NGC 4889 & 269.4 \\
Z 160-235 & -19.89 & 0(0) & 0.00 &  & NGC 4874 & 392.3 \\
IC 4030 & -19.69 & 9(4) & 0.18 &  & NGC 4889 & 274.0
\enddata
\tablecomments{Systems failing the detection criterion, $Z<3$; we report $1\sigma$ upper limits on 
\SN\ derived from $(\psub{N}{corr}+1\sigma)$.}
\end{deluxetable}

In Section~\ref{sec:Coma_gal_GCSF} we examined the GC specific frequency, relating galaxy GCS count to luminosity, \MV. The main detection results were included in the main text (see Table~\ref{tab:raw_GCSF_data_det}). Here, we include the remaining data for non-detections, outline edge cases in the data and how they were handled. The remaining data is included in Table~\ref{tab:raw_GCSF_data_nondet_app}. Several systems yield formally negative or marginal background-corrected counts owing to low intrinsic GC populations and stochastic background fluctuations. These galaxies are therefore classified as non-detections and excluded from quantitative fits. Their upper limits are, however, generally consistent with strong GC deficits and are discussed qualitatively where relevant.

As part of our robustness diagnostics, we assessed the sensitivity of the results to spatial scale, background estimation, completeness, and sampling variance.

\subsection{Aperture and Background variations}
To check the sensitivity of our tests to variations in the chosen aperture and background annulus, we re-performed the main tests outlined in Section~\ref{sec:Coma_gal_GCSF}, estimating the observed background-corrected GC count, \psub{N}{GC,corr}, for different inner radii and background annuli. While we varied both inner and annulus radii, the ratio between the central measurement radius and both the minimum and maximum annulus radii was maintained at $1.5\psub{R}{in}<\psub{R}{ann}<2.5\psub{R}{in}$. The results of this test are shown in Table~\ref{tab:aperture_variations}.

\begin{deluxetable}{cccccccccc}
\tablecaption{Sensitivity to aperture variation\label{tab:aperture_variations}
}
\tablehead{
\colhead{\psub{R}{in} (\Reff)} &
\colhead{\psub{R}{ann} (\Reff)} &
\colhead{\psub{f}{peng}} &
\colhead{\psub{N}{bulk}} &
\colhead{\psub{N}{det}} &
\colhead{\psub{N}{ul}} &
\colhead{\psub{N}{tail}} &
\colhead{Jaccard (vs.8\Reff)} &
\colhead{median residual (det)} &
\colhead{median residual (ul)}
}
\startdata
6 & 9–15 & 0.263 & 26 & 14 & 12 & 11 & 0.79 & -0.06 & -0.75 \\
8 & 12–20 & 0.327 & 26 & 14 & 12 & 14 & 1.00 & -0.03 & -0.80 \\
10 & 15–25 & 0.368 & 26 & 10 & 16 & 13 & 0.93 & 0.04 & -0.73
\enddata
\end{deluxetable}
Across all aperture/background configurations tested (6, 8, and 10 \Reff), the fitted bulk normalization varies modestly, as expected. The bulk-far (BCG-proximity $>\ang{;;70}$) sample size remains fixed ($N=26$), and the median upper-limit residuals for non-detections remain strongly negative (median $\psub{\delta}{UL} \sim -0.7$ to $-0.8$). The corresponding ``low-\SN'' tail persists in all configurations (11–14 systems; Jaccard overlap with the fiducial $8\Reff$ selection of 0.79—0.93), indicating that the identification of a GC-poor tail is robust to reasonable methodological variations.

A larger normalization shift ($f \sim 0.53$) appears in the $10\Reff$ configuration when the closest non-excluded companions to the BCG NGC 4874 (\eg NGC 4871 and NGC 4873) are included in the fit, likely reflecting background annuli that intersect the extended BCG/ICL GC component. Excluding these systems from the normalization fit restores the consistency shown above, without altering the deficit population.
 
\subsection{Completeness and Depth}
Due to the use of different HST programme data and a wide variety of environments throughout our dataset, we perform diagnostics on completeness around the three major ellipticals to test sensitivity to HST pointing and surface brightness.

\begin{figure}
    \centering
    \includegraphics[width=1\linewidth]{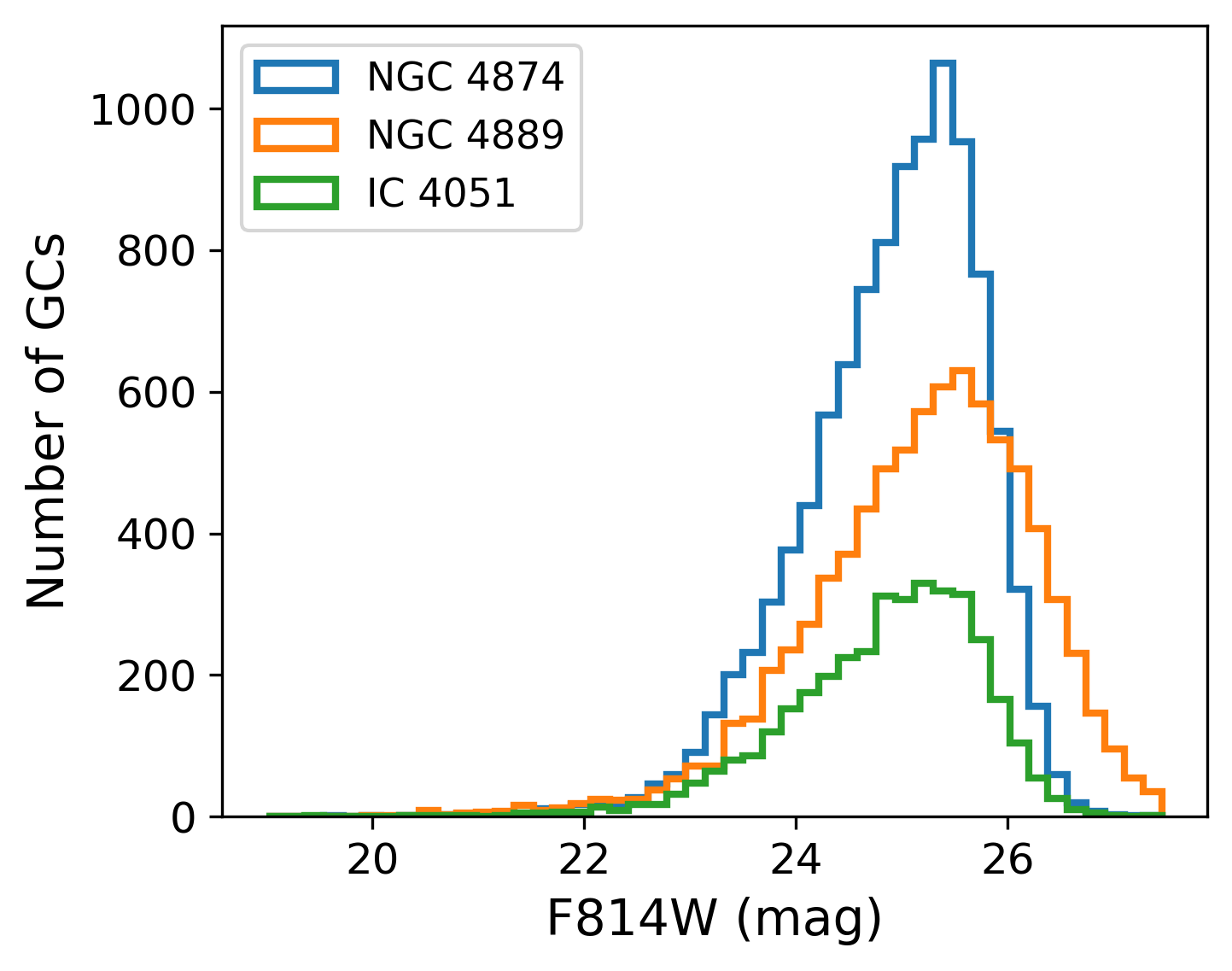}
    \caption[Observed GCLF near giant ellipticals.]{Observed GCLF near giant ellipticals.}
    \label{fig:ellipticals_GCLF}
\end{figure}

We first perform a GCLF diagnostic around each of the three GC-rich ellipticals, which were imaged with different HST programs: NGC 4874 - GO 10861 \citep{Carter_2008}, NGC 4889 - GO 11711 \citep{Cho_2016} and IC 4051 - GO 12918 \citep{Harris_2017b}. The three GCLF histograms, for GCs within $8\Reff$ of the elliptical, are shown overlaid in \cref{fig:ellipticals_GCLF}. We measure the GCLF half-max completeness (knee) points as detailed in Table~\ref{tab:GCLF_knee}. We repeated the knee-completeness analysis separately in inner $(r<2\Reff)$ and outer $(2\Reff<r<8\Reff)$ regions to test for surface-brightness–dependent incompleteness. The derived 50\% knee magnitudes, shown in Table~\ref{tab:GCLF_knee}, agree within $\lesssim0.03$ mag between radial bins for all three galaxies, indicating that crowding and galaxy light do not significantly affect the faint-end detection limit in our sample. However, NGC 4889 consistently exhibits a fainter knee magnitude than the other systems in both radial regimes, reflecting its greater photometric depth.

\begin{deluxetable}{lccccc}
\tablecaption{GCLF 50\% knee-completeness point for GC-rich giant ellipticals\label{tab:GCLF_knee}
}
\tablehead{
\colhead{galaxy} &
\multicolumn{3}{c}{GCLF 50\% knee (mag)} \\
& \colhead{$r\leq~8\Reff$} &
\colhead{$r\leq~2\Reff$} &
\colhead{$2\Reff<r\leq8\Reff$} &
}
\startdata
NGC 4874 & $26.09\pm{0.01}$ & $26.10\pm{0.02}$ & $26.07\pm{0.01}$ \\
NGC 4889 & $26.52\pm{0.02}$ & $26.63\pm{0.05}$ & $26.65\pm{0.03}$ \\
IC 4051 &  $26.08\pm{0.01}$ & $26.04\pm{0.03}$ & $26.07\pm{0.02}$ \\
\enddata
\end{deluxetable}

To assess the impact of this on our GC specific frequency measurements, we repeat the analysis including only a conservative bright magnitude cut at $<25.8$~mag (\ie a nominal 10\% brighter than the 50\% completeness-knee). We compare the vertical normalization factor to the Peng-Virgo curve, between the full data set and the bright cut data, as detailed in Table~\ref{tab:vertical_norm_comp}. The shift in best fit for the vertical normalization required when using the bright-cut data, compared to the full dataset, is minimal under all circumstances and as expected when removing faint objects. However, we note that the structure and relative ordering of the result is unchanged, thus removing the faintest GCs has minimal effect on the normalization. This demonstrates that the main result is not driven by completeness systematics.

\begin{deluxetable}{lccl}
\tablecaption{Comparison between normalization factors with varying test configurations\label{tab:vertical_norm_comp}
}
\tablehead{
\colhead{factor ($f$)} &
\colhead{Original} &
\colhead{Bright-cut} &
\colhead{Definition} 
}
\startdata
all     & $\sim0.87$ & 0.82 & All detections, weighted\\ 
clean   & $\sim0.93$ & 0.84 & Excl. BCGs + edge proximal\\ 
typical & $\sim0.32$ & 0.29 & Also excl. IC 4051\\ 
\enddata
\tablecomments{The bright-cut limits to objects $<25.8$~mag.}
\end{deluxetable}

\subsection{Bootstrap and Uncertainty Propagation}
Bootstrap confidence intervals were computed for the fiducial ($8\Reff$) typical-detection sample to assess the sensitivity of the inferred intrinsic dispersion to sampling noise. For each configuration, 5000 bootstrap resamples were generated, and the robust scatter in $\log_{10}\SN$ was estimated using a median absolute deviation (MAD)–based estimator. The fiducial configuration yields $\sigma_{\log_{10}\SN}=0.19$ dex, with a $16-84\%$ bootstrap interval of $0.12-0.29$ dex ($95\%$: $0.07-0.35$ dex). Comparable values are obtained across reasonable aperture variations. These results are consistent with the intrinsic scatter reported by \citet{Peng_2008}, and indicate that the measured dispersion is not driven by sampling noise or configuration choice.

\subsection{Edge and Coverage Effects}
We recognize that galaxies located close to the edge of HST pointings may have lower quality data associated with them. This is especially true when measuring background subtraction using an annulus which may be more adversely affected by edge proximity than the inner GCS radius. We manually identified the brighter galaxies most likely to be affected, flagged them in the plots, and tested the sensitivity of the results to their exclusion (see Table~\ref{tab:vertical_norm_comp}). We see little impact on the scatter or normalization factors from the inclusion of these galaxies.

We noted the galaxy LEDA~44708 lies only \ang{;;26.8} from NGC~4889 (and \ang{;;40} from NGC~4886), so its apparent GC counts and inferred \SN\ are dominated by the surrounding BCG/intracluster population; we therefore flag it as BCG-proximal and exclude it from host-based fits.

\bibliography{references}{}
\bibliographystyle{aasjournalv7}



\end{document}

%% file: macros.tex
\newcommand{\psub}[2]{\ensuremath{#1_{\mathrm{#2}}}}
\newcommand{\vsub}[2]{\ensuremath{#1_{#2}}}

\newcommand{\eg}{e.g.,\ } 
\newcommand{\ie}{i.e.,\ } 

\newcommand{\Reff}{\ensuremath{\psub{R}{e}}}
\newcommand{\ReffGCS}{\ensuremath{\psub{R}{e,GCS}}}
\newcommand{\SN}{\ensuremath{\psub{S}{N}}}
\newcommand{\MV}{\ensuremath{\vsub{M}{V}}}

\newcommand{\asymcomment}[1]{%
	\parbox[t]{5.2cm}{\raggedright #1}%
}

\newcommand{\stkout}[1]{\ifmmode\text{\sout{\ensuremath{#1}}}\else\sout{#1}\fi}

\DeclareUnicodeCharacter{2609}{\odot}
\DeclareUnicodeCharacter{2605}{\star}

\let\sun\odot

\newcommand*\solarmass{\si{\solarmass}}
\newcommand*\solarluminosity{\si{\solarluminosity}}
\DeclareSIUnit\lightspeed{$c$}
\DeclareSIUnit\rydberg{Ry}
\DeclareSIUnit\erg{erg}
\DeclareSIUnit\magnitude{mag}
\DeclareSIUnit\mag{mag}
\DeclareSIUnit\jansky{Jy}
\DeclareSIUnit\gauss{G}
\DeclareSIUnit\h{$h$}
\DeclareSIUnit\hseven{$h$_7}
\DeclareSIUnit\parsec{pc}
\DeclareSIUnit\year{yr}
\DeclareSIUnit\solarluminosity{\ensuremath{L_\sun}}
\DeclareSIUnit\solarmass{\ensuremath{M_\sun}}
\DeclareSIUnit\solarmassinenergy{\ensuremath{M_\sun|c^2}}
\DeclareSIUnit\solarradius{\ensuremath{R_\sun}}
\DeclareSIUnit\astronomicalunit{AU}
\DeclareSIUnit\clight{\ensuremath c}

\newcommand{\IAIFI}{The NSF AI Institute for Artificial Intelligence and Fundamental Interactions}
\newcommand{\CfA}{Center for Astrophysics $|$ Harvard \& Smithsonian, Cambridge, MA 02138, USA}
\newcommand{\MIT}{Department of Physics, Massachusetts Institute of Technology, Cambridge, MA 02139, USA}


%% file: variables.tex
\def \RTPVarCSSTotal{23,351\xspace} 
\def \RTPVarGCTotal{22,828\xspace} 
\def \RTPVarUCDTotal{523\xspace} 